\documentclass[draftclsnofoot,journal,11pt,onecolumn]{IEEEtran}

\usepackage[utf8]{inputenc} 
\usepackage[T1]{fontenc}

\usepackage{amsfonts, amsmath, amssymb}
\usepackage{bm, bbm}
\usepackage{mathtools}
\usepackage{commath}
\usepackage{stmaryrd}
\usepackage{faktor}

\usepackage[ruled]{algorithm2e}

\usepackage{graphicx}
\usepackage[caption=false, font=footnotesize]{subfig}

\usepackage[shortlabels]{enumitem}

\usepackage{array}
\usepackage{textcomp}
\usepackage{stfloats}
\usepackage{url}
\usepackage{verbatim}
\usepackage{cite}

\usepackage{color}

\newcommand{\suchthat}{\; \colon \;}
\newcommand{\given}{\,\vert\,}
\newcommand{\mgiven}{\;\middle\vert\;}

\renewcommand{\hat}{\widehat}
\renewcommand{\tilde}{\widetilde}

\renewcommand{\d}{\mathrm{d}}

\newcommand{\error}{\mathrm{error}}

\newcommand{\KT}{\mathrm{KT}}
\newcommand{\ML}{\mathrm{ML}}

\newcommand{\MLE}{\mathrm{MLE}}

\newcommand{\1}{\mathbbm{1}}

\newcommand{\E}{\mathbb{E}}

\newcommand{\N}{\mathbb{N}}

\newcommand{\R}{\mathbb{R}}

\newcommand{\Pbb}{\mathbb{P}}

\newcommand{\Acal}{\mathcal{A}}

\newcommand{\Ccal}{\mathcal{C}}

\newcommand{\Ecal}{\mathcal{E}}
\newcommand{\Fcal}{\mathcal{F}}

\newcommand{\Mcal}{\mathcal{M}}
\newcommand{\Pcal}{\mathcal{P}}

\newcommand{\Scal}{\mathcal{S}}
\newcommand{\Ucal}{\mathcal{U}}
\newcommand{\Tcal}{\mathcal{T}}

\newcommand{\sv}{\pmb{s}}

\newcommand{\xv}{\pmb{x}}
\newcommand{\yv}{\pmb{y}}
\newcommand{\zv}{\pmb{z}}

\newcommand{\Sv}{\pmb{S}}
\newcommand{\Xv}{\pmb{X}}
\newcommand{\Yv}{\pmb{Y}}
\newcommand{\Zv}{\pmb{Z}}

\DeclareMathOperator{\Var}{Var}

\newtheorem{theorem}{Theorem}
\newtheorem{lemma}{Lemma}

\newtheorem{remark}{Remark}

\usepackage{mathtools}
\makeatletter
\newcommand{\vast}{\bBigg@{3}}
\newcommand{\Vast}{\bBigg@{4}}

\makeatother
\begin{document}

\title{Universal Decoding over Finite-State\\ Additive Channels via Noise Guessing}

\author{Henrique~K.~Miyamoto,~\IEEEmembership{Graduate Student Member,~IEEE}
		and
		Sheng~Yang,~\IEEEmembership{Member,~IEEE}%
	\thanks{The authors are with Université Paris-Saclay, CNRS, CentraleSupélec, Laboratoire des Signaux et Systèmes~(L2S), 91190, Gif-sur-Yvette, France (e-mail: henrique.miyamoto@centralesupelec.fr; sheng.yang@centralesupelec.fr).}%
	\thanks{A preliminary version of this work has been accepted to the 2025 IEEE International Workshop on Information Theory (ITW~2025).}%
}



\maketitle

\begin{abstract}
	We study universal decoding over unknown discrete additive channels determined by a finite-state (unifilar) random process.
	Aiming at low-complexity decoders, we study variants of noise-guessing decoders that use estimators for the probability of a noise sequence when the actual channel law
	is unknown.
	A deterministic version produces noise sequences in a fixed order, and a new randomised version draws them at random, until finding a sequence that, subtracted from the received sequence, results in a valid codeword.
	We show that both strategies are random-coding universal (i.e. have the same random-coding error exponent as the optimal maximum likelihood decoding), and derive upper bounds for their complexity.
	Numerical examples in additive Markov channels illustrate the proposed methods' performance, showing that they consistently outperform a more usual training-based strategy.
\end{abstract}


\section{Introduction} \label{sec:introduction}

Universal decoders are those that, despite not knowing the specific probability law of the channel in use (but only the family to which it belongs), have a random-coding error exponent that is asymptotically the same as that of the optimal maximum likelihood~(ML) decoder tuned for the channel in question~\cite{lapidoth1998b}. This is of practical interest, since, in reality, one may not know the statistics of the channel over which communication takes place, and the usual estimate-and-decode strategy comes with an inherent trade-off between error probability and communication rate~\cite{feder1998}. Indeed, if the channel law is unknown, one cannot use the optimal ML decoder, or any other decoder that relies on the channel parameters.

While universal decoders have been studied in a variety of scenarios (e.g.~\cite{goppa1975a,csiszar2011,ziv1985,merhav1993,lapidoth1998a,feder1998,feder2002,shayevitz2005,merhav2013,huleihel2015,averbuch2018,tamir2022,miyamoto2024} and references therein), a key difficulty in practically implementing them lies in the computational complexity.
In fact, they generally require evaluating the decoding metric of each codeword in the codebook, which, although being the same complexity of brute-force ML decoding, becomes prohibitive for large codebooks.
Notable exceptions to that are the modified stack algorithms used with convolutional codes over unknown channels~\cite{ziv1985,lapidoth1999}, and the decoders for frequency-selective fading channels studied in~\cite{shayevitz2005}.

An alternative implementation of ML decoding in discrete additive channels, based on noise guessing, was proposed in~\cite{duffy2019} and named guessing random additive noise decoding~(GRAND)\footnote{
	In some contexts, GRAND (and other decoders) are referred to as `universal' for not being specific to a given codebook structure. Here, we follow the established nomenclature in the information theory literature~\cite{goppa1975a,csiszar2011,ziv1985,merhav1993,lapidoth1998a,feder1998,feder2002,shayevitz2005,merhav2013,huleihel2015,averbuch2018,tamir2022,miyamoto2024} and reserve the expression `universal decoder' for a decoder that is essentially optimal without knowing the channel law (i.e. the universality aspect is with respect to the channel distribution, and not to the code).
}. The idea is to sequentially query noise sequences until finding one that, subtracted from the received sequence, corresponds to a valid codeword. A main advantage of this approach is that its average complexity can be much lower than that of brute-force ML decoding, notably, in low-noise channels or with high-rate codes. The optimal order to query noise sequences is that of decreasing noise sequence probabilities, which is usually assumed to be known or estimated~\cite{duffy2019,an2022}.
On the other hand, a version that does not require knowledge of the noise distribution, and can achieve Gallager's error exponent in memoryless additive channels has been recently studied in \cite{jouhed2024,tan2025}. We follow this direction and take some steps further towards low-complexity universal decoding.

Specifically, inspired by universal compression, we consider universal decoders based on maximising and weighting estimators for the noise distribution, when the actual channel law is unknown. We extend decoding by \emph{deterministic noise guessing}, in which noise sequences are queried in a fixed order, beyond memoryless channels; and we introduce decoding by \emph{randomised noise guessing}, in which noise sequences are randomly drawn---in the same spirit of universal randomised guessing~\cite{merhav2020}, and somewhat analogously to stochastic decoders~\cite{scarlett2015,merhav2017}.
Leveraging well-known techniques based on the method of types for finite-state distributions~\cite{csiszar1998}, we study these two decoding strategies on unifilar additive channels (which include memoryless and Markov additive channels, for instance). We show that these decoders are random-coding universal (i.e. have the same random-coding error exponent as the optimal ML decoding), and derive upper bounds on their average complexity (i.e. number of guesses needed to find a codeword), which turn out to match the asymptotic complexity of GRAND with known channel law~\cite{duffy2019}.
In particular, we establish that decoding by randomised noise guessing has both error probability and complexity exponents that are asymptotically the same as those of the deterministic counterpart, while having potential advantages for practical implementation, as discussed next.
We also remark that our upper bound results on the decoding `redundancy' (difference of error exponents) and complexity are non-asymptotic, thus providing some insight into the rate of convergence towards their asymptotic values.

We propose practical schemes for both the deterministic and randomised strategies, and show that the latter can provide a very simple implementation in which queries can be generated with linear complexity in the block-length~$n$ and that can be completely parallelised without the need of coordination between the parallel units. This can be particularly interesting when the channel memory grows large, as the complexity of the deterministic strategy, albeit polynomial in~$n$, scales exponentially with the number of states.
The schemes are evaluated with emphasis on short packet length scenarios, which can be of special relevance in applications that have low-latency requirements. Numerical results show that the universal strategies consistently outperform a more usual training-based approach that first estimates the channel and then decode based on the estimated parameters. In fact, it is already known that the training-based strategy does not yield, in general, a universal decoder~\cite{feder1998}.

The remainder of this paper is organised as follows. Section~\ref{sec:preliminaries} formalises the problem, sets definitions and preliminary results. The decoding by deterministic noise guessing strategy is presented and analysed in Section~\ref{sec:deterministic-guessing}, and the randomised counterpart in Section~\ref{sec:randomised-guessing}. Numerical results are presented in Section~\ref{sec:numerical-results}, and Section~\ref{sec:conclusion} concludes the paper.

\paragraph*{Notation} We denote scalars by italic letters (e.g.~$x$) and vectors by italic, bold face letters (e.g.~$\xv$); random variables by upper case letters (e.g. $X$, $\Xv$) and the corresponding realisations by lower case letters (e.g. $x$, $\xv$). Logarithms are to base~$2$. We denote $\1_{\Scal}(x)$ the indicator function of set $\Scal$, which takes value $1$, if $x \in \Scal$, and $0$ otherwise.

\section{Preliminaries} \label{sec:preliminaries}

\subsection{Universal Decoding} \label{subsec:universal-decoding}

Consider a discrete additive channel over the finite alphabet $\Acal$, which we identify with the set $\{1, \dots, |\Acal|\}$. Upon $n$ uses, the transmitter sends the sequence $\Xv \coloneqq X_1^n \coloneqq X_1 \cdots X_n \in \Acal^n$ and the receiver observes $\Yv \coloneqq Y_1^n \in \Acal^n$, given by
\begin{equation}
	\Yv = \Xv + \Zv,
\end{equation}
where addition is modulo-$|\Acal|$, and $\Zv \coloneqq Z_1^n \in \Acal^n$ is a noise sequence independent of $\Xv$. Let
\begin{equation*}
	\Fcal \coloneqq \left\{ P_{\theta} \suchthat \theta \in \Theta \subseteq \R^d \right\}
\end{equation*}
be a parametric family of random processes over $\Acal$, of fixed dimension~$d$. The distribution $P_{\Zv} \coloneqq P_{\Zv;\theta}$ of $\Zv \in \Acal^n$ is the restriction of $P_\theta \in \Fcal$ to $n$-length sequences. The channel is then characterised by the channel law
\begin{equation*}
	P_{\Yv \given \Xv}\left(\yv \mgiven \xv\right) = P_{\Zv}(\yv - \xv).
\end{equation*}

A codebook $\Ccal \coloneqq \left( \xv_1, \cdots, \xv_M \right) \subseteq \Acal^n$ of rate $R \coloneqq \left(\log M\right)/n$ and block-length~$n$ is employed, so that the transmitter conveys message $m \in \Mcal \coloneqq \{1, \dots, M\}$ by transmitting codeword $\xv_m \in \Ccal$. The decoder implements a decoding rule (deterministic or randomised) $\varphi_n \colon \Acal^n \to \Mcal$ to recover the transmitted message. We assume equiprobable messages, for which the optimal decoding rule is the ML rule $\varphi_{\ML}(\yv) = \arg\max_{m\in\Mcal} P_{\Zv}(\yv-\xv_m)$, with the convention that ties resolved at random among codewords realising the maximum.

In the following, we consider uniform random coding over $\Acal^n$: codewords are chosen independently and uniformly over $\Acal^n$, that is, with distribution $P_{\Xv}(\xv) = \frac{1}{|\Acal|^n}$. We denote $P_{\theta,\varphi}(\error)$ the average probability of error (over messages and random codes) of decoder~$\varphi$ when used in channel~$\theta$. Sometimes we replace the subscript $\varphi$, as in $P_{\theta,\ML} \coloneqq P_{\theta,\varphi_{\ML}}$, and as indicated later in the text.
Not without some abuse\footnote{
	In analogy to the context of universal source coding (e.g.~\cite{davisson1973}).
}, we call \emph{redundancy} of a decoder $\varphi_n$ the difference in the random-coding error exponent in comparison to the optimal ML decoder for channel $\theta$, that is,
\begin{equation}
	R_{n} \left( \varphi_n, \theta \right) \coloneqq \frac{1}{n}\log \left( \frac{P_{\theta,\varphi_n}(\error)}{P_{\theta,\ML}(\error)} \right).
\end{equation}
Following \cite[Definition~3]{feder1998}, we say that a sequence of decoders $(\varphi_n)_{n\in\N}$ is \emph{random-coding (weakly) universal} for the family of channels $\Fcal = \{ P_{\theta} \suchthat \theta \in \Theta \}$, if
\begin{equation}
	\lim_{n\to\infty} R_{n}\left( \varphi_n, \theta \right) = 0, \quad \forall \theta \in \Theta,
\end{equation}
and \emph{random-coding strongly universal}, if
\begin{equation}
	\lim_{n\to\infty} \sup_{\theta \in \Theta} R_{n}\left( \varphi_n, \theta \right) = 0.
\end{equation}

\subsection{Channel Model}

We will consider the family~$\Fcal$ of additive unifilar channels~\cite{ash1990,weinberger1992}, that is, additive channels in which the noise sequence is generated by a finite-state process with deterministic transitions~\cite{ziv1985}.
These are characterised by a finite set of states $\Scal$, identified with $\left\{ 1, \dots, |\Scal| \right\}$; probability measures $P_{Z \given S}\left( z \given s \right)$, defined for $z\in\Acal$ and $s\in\Scal$; and a next-state function $f \colon \Acal \times \Scal \to \Scal$. The latter will be assumed to be such that the restriction map $f(\cdot,s) \colon \Acal \to \Scal$ is injective for each $s\in\Scal$ so that, given the initial state $s_0 \in \Scal$, the sequence of states $\sv \coloneqq s_1^n \in \Scal^n$ is completely determined by the sequence of symbols $\zv \in \Acal^n$, and vice-versa, through the relation $s_i = f\left(z_i,s_{i-1}\right)$, $1 \le i \le n$.
For the initial state $s_0 \in \Scal$, the probability of a sequence $\zv\in\Acal^n$ is
\begin{equation} \label{eq:unifilar-distribution}
	P_{\Zv}(\zv \given s_0)
	\coloneqq P_{\Zv;\theta}(\zv \given s_0)
	= \prod_{i=1}^{n} P_{Z \given S}\left(z_{i} \given s_{i-1}\right).
\end{equation}

Denote $\Delta^{n-1} \coloneqq \left\{ (p_1, \dots, p_n) \in \left[0,1\right]^n \suchthat \sum_{i=1}^{n} p_i = 1 \right\}$ the $(n-1)$-simplex. Given the set of states~$\Scal$ and next-state function~$f$, the distribution $P_{\Zv;\theta}$ can be parametrised by a stochastic matrix $\theta \coloneqq \left(\xi_{s}(z)\right)_{z\in\Acal,\,s\in\Scal} \in \Delta^{(|\Acal|-1)|\Scal|}$, by taking $P_{Z \given S}(z \given s) = \xi_{z}(s)$. This model induces a Markov chain over states $\Scal$ with transition distribution $P_{S' \given S}( s' \given s ) \coloneqq \sum_{z\in\Acal} P_{Z \given S}(z \given s) \1_{\{f(z,s)\}}(s')$. We will later restrict ourselves to the family $\Fcal^{\star} \subset \Fcal$ of unifilar distributions whose induced Markov chain is stationary and irreducible.

Note that the model defined by~\eqref{eq:unifilar-distribution} includes additive channels whose noise sequence is generated by memoryless laws ($|\Scal|=1$), $k$-th order Markov processes ($\Scal = \Acal^k$), and, more generally, finite-state context trees (also known as FSMX in~\cite{rissanen1986}).

\subsection{Method of Types} \label{subsec:method-of-types}

We use the method of types for finite-state sequences~\cite[Section~VII]{csiszar1998}. Given initial state $s_0 \in \Scal$, consider the equivalence relation ${\zv \sim \zv'}$ defined by $\forall \theta \in \Theta,\ P_{\Zv; \theta}(\zv \given s_0) = P_{\Zv, \theta}(\zv' \given s_0)$; when $\zv \sim \zv'$ the sequences $\zv, \zv' \in \Acal^n$ are said to have the same \emph{type}.
Each type induces a distribution, namely, the one parametrised by the maximum likelihood estimator~(MLE) $\hat{\theta}_{\MLE}(\zv)$, and we denote $\hat{P}_{\zv} \coloneqq P_{\hat{\theta}_{\MLE}(\zv)} \in \Fcal$.
Denote $\Pcal_n(\Acal) \coloneqq \big\{ \hat{P}_{\zv} \suchthat \zv \in \Acal^n \big\} \subseteq \Fcal$ the set of all such distributions (which can be identified with the quotient set $\faktor{\Acal^n}{\sim}$), and $\Tcal_n(\hat{P}_{\zv}) \coloneqq \left\{ \zv' \in \Acal^n \suchthat \zv \sim \zv' \right\}$ the equivalence class (\emph{type class}) of~$\zv \in \Acal^n$.

Given an initial state\footnote{
	The dependency on the initial state will be henceforth omitted and should be understood.
} $s_0 \in \Scal$, finite-state types are characterised by the counts $a_{\zv}(z,s) \coloneqq a_{\zv,s_0}(z,s) \coloneqq \sum_{i=1}^{n} \1_{\{(z,s)\}}\left(z_i,s_{i-1}\right) $ and $a_{\sv}(s) \coloneqq \sum_{z\in\Acal} a_{\zv}(z,s)$. The counts define the empirical distributions
\begin{equation*}
	\hat{P}_{\zv}\left(z,s\right) \coloneqq \frac{a_{\zv}(z,s)}{n}
\end{equation*}
and
\begin{equation*}
	\hat{P}_{\zv}\left(z \given s\right) \coloneqq
	\begin{cases}
		\frac{a_{\zv}(z,s)}{a_{\zv}(s)}, & \text{if } a_{\zv}(s) \neq 0,\\
		0, & \text{otherwise}.
	\end{cases}
\end{equation*}
Note that the MLE $\hat{\theta}$ is composed by the parameters $\hat{\xi}_s(z) \coloneqq \hat{P}_{\zv}\left( z \given s \right)$.

Defining the quantities
\begin{equation}
	{H}(\hat{P}_{\zv})
	\coloneqq \sum_{(z,s)\in\Acal\times\Scal} \hat{P}_{\zv}(z,s) \log \left(\frac{1}{\hat{P}_{\zv}(z \given s)}\right)
\end{equation}
and
\begin{equation}
	{D}(\hat{P}_{\zv} \| P_{\Zv;\theta})
	\coloneqq \sum_{(z,s)\in\Acal\times\Scal} \hat{P}_{\zv}(z,s) \log \left(\frac{\hat{P}_{\zv}(z \given s)}{P_{Z \given S}(z \given s)}\right),
\end{equation}
we can write
\begin{equation}
	{P}_{\Zv;\theta}(\zv)
	= 2^{-n\left({H}(\hat{P}_{\zv}) + {D}(\hat{P}_{\zv} \| P_{\Zv;\theta}) \right)}.
\end{equation}

The next lemma summarises well-known results on finite-state types (see e.g. \cite{davisson1981,csiszar1998,martin2012}).

\begin{lemma} \label{lemma:method-of-types}
	We have
	\begin{equation} \label{eq:number-types}
		\left|\Pcal_n\right| \le 2^{n\zeta(n)},
	\end{equation}
	with $\zeta(n) \coloneqq |\Acal||\Scal| \frac{\log (n+1)}{n}$, and, for every $\zv\in\Acal^n$,
	\begin{equation} \label{eq:size-type-class}
		2^{n\left(H(\hat{P}_{\zv}) - \delta(n)\right)}
		\le \left| \Tcal_n(\hat{P}_{\zv}) \right|
		\le 2^{nH(\hat{P}_{\zv})},
	\end{equation}
	with $\delta(n) = O\left(\frac{\log n}{n}\right)$.
\end{lemma}

\subsection{Maximising and Weighting}

When the parameter~$\theta$ of the actual distribution $P_{\Zv;\theta}$ is unknown, but we want to estimate the probability of a given sequence $\zv \in \Acal^n$, two strategies will be considered, in an analogous fashion to universal compression techniques~\cite{davisson1973}.
The first strategy is to use the \emph{maximising} distribution:
\begin{equation} \label{eq:maximising-estimator}
	P_{m}(\zv)
	\coloneqq
	\frac{\sup_{\theta\in\Theta} P_{\Zv;\theta}(\zv)}{\sum_{\zv'\in\Acal^n}\sup_{\theta\in\Theta} P_{\Zv;\theta}(\zv')}
	= \frac{\hat{P}_{\zv}(\zv)}{\sum_{\zv'\in\Acal^n}\hat{P}_{\zv'}(\zv')},
\end{equation}
which corresponds to Shtarkov's normalised maximum likelihood (NML) distribution. 
The second is to use a \emph{weighting} distribution:
\begin{equation} \label{eq:weighting-estimator}
	P_{w}(\zv)
	\coloneqq \E_{\theta}\left[ P_{\Zv;\theta}(\zv) \right]
	= \int_{\Theta} P_{\Zv;\theta}(\zv)\,\d\pi(\theta),
\end{equation}
that is, a Bayesian mixture with respect to some prior distribution $\pi$ on $\Theta$. In particular, weighting with independent Dirichlet priors with hyperparameters $\left( 1/2, \dots, 1/2 \right)$ on each conditional distribution yields the Markov Krichevsky--Trofimov~(KT) distribution
\begin{equation} \label{eq:kt-distribution}
	P_{\KT}\left(\zv \given s_0\right)
	= \frac{ \Gamma \big( \frac{|\Acal|}{2} \big)^{|\Scal|} }%
	{ {\Gamma \big( \frac{1}{2} \big)}^{|\Acal||\Scal|} } \prod_{s\in\Scal} \frac{ \prod_{z \in \Acal} \big( a_{\zv}(z,s) + \frac{1}{2} \big) }{ \Gamma \big( a_{\sv}(s) + \frac{|\Acal|}{2} \big) }.
\end{equation}

The next result states that this weighting distribution is sufficiently close to the type distribution, in an exponential sense (see e.g. \cite[Proposition~2.18]{gassiat2018}).

\begin{lemma} \label{lemma:weighting}
	We have
	\begin{equation} \label{eq:weighting-close-type}
		\max_{\zv \in \Acal^n}
		\left(
		\frac{\hat{P}_{\zv}\left(\zv \given s_0\right)}{P_{\KT}\left(\zv \given s_0\right)}
		\right)
		\le 2^{n\epsilon(n)},
	\end{equation}
	with $\epsilon(n) \coloneqq \frac{|\Scal|(|\Acal|-1)}{2}\frac{\log n}{n} + \frac{2|\Scal|}{n}$.
\end{lemma}

\subsection{Entropy Rates}

We now consider the subfamily $\Fcal^{\star} \subset \Fcal$ of unifilar distributions characterised by stationary, irreducible Markov chains. Let $P,Q \in \Fcal^{\star}$, and denote $P_S$ the unique stationary distribution on states induced by $P$. We write
\begin{equation} \label{eq:entropy-rate}
	{H}(P)
	\coloneqq \sum_{(z,s)\in\Acal\times\Scal} P_S(s) P_{Z \given S}(z \given s) \log \left(\frac{1}{P_{Z \given S}(z \given s)}\right),
\end{equation}
the entropy rate, and
\begin{equation} \label{eq:divergence-rate}
	{D}(P \| Q)
	\coloneqq \sum_{(z,s)\in\Acal\times\Scal} P_S(s) P_{Z\given S}(z \given s) \log \left(\frac{P_{Z \given S}(z \given s)}{Q_{Z \given S}(z \given s)}\right),
\end{equation}
the divergence rate associated to these distributions.

Recall that the Rényi entropy of order $\alpha>0$, $\alpha \neq 1$, of a process $P_{\Zv;\theta}$ is $H_{\alpha}(P_{\Zv;\theta}) \coloneqq {\left(1-\alpha\right)}^{-1} \log \left( \sum_{\zv\in\Acal^n} P_{\Zv;\theta}(\zv)^{\alpha} \right)$. The corresponding Rényi entropy rate of $P_\theta \in \Fcal$ is ${H}_{\alpha}(P_{\theta}) \coloneqq \lim_{n\to\infty} n^{-1} H_{\alpha}(P_{\Zv;\theta})$, if the limit exists.
The next result gives a variational characterisation of Rényi entropy rate for unifilar distributions defined by stationary, irreducible Markov chains (see \cite[Example~3]{hanawal2011}).
\begin{lemma} \label{lemma:renyi-entropy}
	For every $P_{\theta} \in \Fcal^{\star}$ and $0 < \alpha < 1$, we have
	\begin{equation} \label{eq:renyi-entropy-variational}
		{H}_{\alpha}(P_\theta) = \max_{Q \in \Fcal^{\star}} \left( {H}(Q) - \frac{\alpha}{1-\alpha} {D}(Q \| P_\theta) \right).
	\end{equation}
\end{lemma}

Note that the distributions in $\Pcal_n(\Acal) \subset \Fcal$ determined by the type are not necessarily stationary and irreducible. Nevertheless, this is not too serious for our purposes, because we can have the following bound.

\begin{lemma} \label{lemma:bound-variational-characterisation}
	For every $P_{\theta} \in \Fcal^{\star}$ and $0 < \alpha < 1$, we have
	\begin{align} \label{eq:renyi-entropy-variational-bound}
		\max_{\hat{P}_Z \in \Pcal_n(\Acal)} \left( {H}(\hat{P}_Z) - \frac{\alpha}{1-\alpha} {D}(\hat{P}_Z \| P_\theta) \right) 
		\le
		{H}_{\alpha}(P_\theta)
		+ \sigma_{\theta}(n),
	\end{align}
	where $\sigma_{\theta}(n) \coloneqq \frac{|\Scal|}{n\left(1-\alpha\right)} \left( |\Acal| \log e + \log |\Acal| - \alpha |\Acal| \log P^*_\theta \right)$, and $P^*_{\theta} \coloneqq \min \left\{ P_{\theta}(z \given s) \colon P_{\theta}(z \given s) > 0\right\}$.
\end{lemma}
\begin{IEEEproof}
	See Appendix~\ref{sec:appendix-proof-lemma-variational}.
\end{IEEEproof}

\section{Decoding by Deterministic Noise Guessing} \label{sec:deterministic-guessing}

Directly implementing a decoder of the form
\begin{equation} \label{eq:codeword-evaluation}
	\varphi_u(\yv) = \arg\max_{m\in\Mcal} u\left(\yv - \xv_m\right)
\end{equation}
requires computing the metric\footnote{
	Following established use, we apply the word `metric', even if this function is not a distance, in the sense of metric spaces.
}~$u \colon \Acal^n \to \R$ of all codewords $\xv_m \in \Ccal$ in the codebook, and selecting the maximising one, a procedure that rapidly becomes unfeasible for large codebooks.
Indeed, a straightforward implementation of~\eqref{eq:codeword-evaluation} would require $O(Mn)$ computations and $O(Mn)$ storage. If the code is structured, e.g. linear, the storage requirement can be alleviated, but, anyhow, for a fixed rate~$R$, the exponential scaling $M=2^{nR}$ becomes prohibitive as the block-length~$n$ grows.

Instead of that, another possible strategy is decoding by deterministic noise guessing, also known as guessing random additive noise decoding~(GRAND) in~\cite{duffy2019} (see also~\cite{jouhed2024}). The idea is to rank noise sequences $\zv \in \Acal^n$, and query whether $\yv - \zv \in \Ccal$, in the ranking order, until a codeword is found\footnote{
	Other stopping criteria could be considered, as in guessing with abandonment, studied in~\cite{duffy2019,tan2025}, but we do not analyse those here.
}.
Denote $G_u \colon \Acal^n \to \{ 1, \dots, |\Acal|^n \}$ a \emph{ranking function} (a bijection) that ranks sequences according to function $u$, i.e. such that ${G_u(\zv) < G_u(\zv')} \implies {u(\zv) \ge u(\zv')}$. In the following, we will be interested in \emph{maximising deterministic guessing decoding} with the guessing function~$G_m$, based on the function $u_m(\zv) \coloneqq P_m(\zv)$, and \emph{weighting deterministic guessing decoding}, with the guessing function~$G_w$, based on the function $u_w(\zv) \coloneqq P_{w}(\zv)$.

\begin{remark} \label{remark:codeword-evaluation}
	Note that guessing decoding with function $G_u$ is equivalent to decoding with~\eqref{eq:codeword-evaluation} using the same function~$u$. Indeed, suppose that deterministic noise decoding decodes sequence $\yv \in \Acal^n$ as $\hat{m}$; this means that $G_u(\yv - \xv_{\hat{m}}) < G_u(\yv - \xv_{m'})$ for all $m' \neq \hat{m}$, implying that $u(\yv - \xv_{\hat{m}}) \ge u(\yv - \xv_{m'})$. In particular, guessing decoding with $G^{\star}$, the ranking function based on $u(\zv) = P_{\Zv}(\zv)$, is equivalent to ML decoding~\cite{duffy2019}. Guessing with the maximising distribution $u_m(\zv) = P_m(\zv)$ corresponds to the generalised likelihood ratio test~(GLRT) decoder~\cite[p.~2166]{lapidoth1998b}; but, unlike in the case of general (not necessarily additive) discrete memoryless channels, here, this decoder is not a maximum mutual information or minimum conditional entropy decoder, but rather a minimum (noise) entropy decoder, that is, one that looks for a codeword~$\xv_m$ that minimises the quantity $H(\hat{P}_{\yv-\xv_m})$.
\end{remark}

\subsection{Decoding Scheme} \label{subsec:deterministic-guessing-scheme}

There are two main steps involved in deterministic noise guessing, as in GRAND~\cite{duffy2019}. In the following, we describe the proposed scheme in the case of additive unifilar channels, particularly for maximising and weighting deterministic guessing.

\paragraph*{Step~1: Generate the guesses}

If $u(\zv)$ only depends on the sequence $\zv$ via its type $\hat{P}_{\zv}$, generating the guesses can be done type-wise: first, compute $u(\zv)$ for one sequence of each type $\hat{P}_{\zv}$; then, for each type, following decreasing order of $u$-value, generate (i.e. enumerate) the sequences in the type class of that type.
Computing the $u$-values for all types requires $O\big(n^{|\Acal||\Scal|}\big)$ computations and $O\big(n^{|\Acal||\Scal|}\big)$ storage, and additional $O\big( |\Acal||\Scal| n^{|\Acal||\Scal|} \log n \big)$ computations and $O(1)$ storage to sort them.
While enumeration of sequences is quite straightforward in the memoryless case, efficient enumeration algorithms similar to that of~\cite{cover1973} can be applied when the channel has memory or states. A pseudocode\footnote{
	This pseudocode considers the distribution on the state sequences and generates the state sequence $s_0^n \in \Scal^{n+1}$, but the corresponding symbol sequence $z_1^n \in \Acal^n$ can be easily obtained with the next-state function~$f$. The sequence $s_1^{0}$ should be understood as the empty sequence.
} is provided in Algorithm~\ref{algo:enumeration}, which only depends on the size of finite-state type-class, cf.~\cite[Equation~(1)]{martin2012}.

\begin{algorithm}
	\caption{Sequence enumeration in $\Tcal_n(\hat{P}_S)$} \label{algo:enumeration}
	\KwIn{Index $i \in \big\{ 1,\dots,|\Tcal_n(\hat{P}_S)| \big\}$, type $\hat{P}_{S}$, initial state $s_0$.}
	\KwOut{Sequence $s^n_0 \in \Scal^{n+1}$.}
	\For{$k \in \left\{1,\dots,n\right\}$}%
	{
		$i_k \gets 0$\\
		\For{$s \in \Scal$}%
		{
			Compute $N(s_1^{k-1}s) \coloneqq
			\left| \left\{ \tilde{s}^n_1 \in \Tcal_n(\hat{P}_{S}) \suchthat \tilde{s}^{k}_1 = s_1^{k-1}s \right\} \right|$\\
			\eIf{$i_k + N(s_1^{k-1} s) > i$}
			{
				$s_k \gets s$\\
				$i \gets i-i_k$\\
				\textbf{break} for loop
			}
			{
				$i_k \gets i_k + N(s_1^{k-1} s)$\\		
			}
		}
	}
\end{algorithm}

\paragraph*{Step~2: Query the sequences}

Upon observing $\yv$, the decoder should generate noise sequences $\zv$ (cf. step~1), and query whether $\yv - \zv$ is a codeword until a positive answer is found. Checking whether $\yv-\zv$ is a codeword could require a search among the codewords of an unstructured code, which has storage complexity $O(M)$, but if the code is linear, for instance, this can be done by multiplying the sequence by the parity-check matrix, with both computational and storage complexity $O(n^2)$. The number of queries needed in this step is a random variable and controls the complexity of the decoding scheme.

Note that, in step~1, if the channel is known and the ranking function $G_u = G^\star$ is used, then it corresponds to the algorithm of~\cite{duffy2019}. Generating the sequences in decrease order of probability is straightforward in the memoryless case, and, in the case of Markov noise of order $1$, a procedure was presented in~\cite{an2022}, under assumption of known channel parameters. The proposed implementation of step~1, on the other hand, can be applied in the more general case of unifilar distributions, and with any function~$u$ that only depends on the sequences via their type, including those that do not correspond to the actual noise distribution, as in the case of maximising~$u_m$ and weighting~$u_w$ metrics.

\subsection{Main Results}

In the following, we state the main results concerning both the probability of error (universality) and the complexity of maximising and weighting deterministic guessing. The proofs are presented in the end of the section.
Denote $P_{\theta,G_u}(\error)$ the average probability of error and $Q_{\theta,G_u}(\error)$ the average number of guesses needed to find a codeword with decoding by deterministic noise guessing with ranking function~$G_u$.

\subsubsection{Universality}

The following result upper bounds the redundancy and establishes the universality of both decoding strategies.

\begin{theorem}[Universality] \label{thm:determistic-guessing-universality}
	(a)~The maximising deterministic guessing decoder is random-coding strongly universal for additive unifilar channels, with
	\begin{equation} \label{eq:maximising-determistic-guessing-universality}
		\frac{1}{n}\log\frac{P_{\theta,G_m}(\error)}{P_{\theta,\ML}(\error)}
		\le \frac{1}{n} + \zeta(n) + \delta(n).
	\end{equation}

	(b)~The weighting deterministic guessing decoder is random-coding strongly universal for additive unifilar channels, with
	\begin{equation} \label{eq:weighting-determistic-guessing-universality}
		\frac{1}{n}\log\frac{P_{\theta,G_w}(\error)}{P_{\theta,\ML}(\error)}
		\le \frac{1}{n} + \epsilon(n) + \delta(n).
	\end{equation}
\end{theorem}

Note that both redundancies vanish as $O\left( (\log n)/n \right)$.
Part~(a) of Theorem~\ref{thm:determistic-guessing-universality}, when specialised to memoryless additive channels, strengthens the result of~\cite[Equation~(42)]{jouhed2024}, which shows that a universal version of GRAND (equivalent to maximising deterministic guessing decoder) can asymptotically achieve Gallager's random coding exponent for memoryless channels. Our result refines that by giving an upper bound on the decoder redundancy, and extends it to the weighting decoder and additive unifilar channels.
Furthermore, in view of Remark~\ref{remark:codeword-evaluation}, the results of Theorem~\ref{thm:determistic-guessing-universality} also apply to the corresponding decoders $\varphi_{u_m}$ and $\varphi_{u_w}$.
In the special case of memoryless additive channels, these results are somewhat analogous to~\cite[Lemma~1]{ziv1985} and \cite[Theorem~1]{miyamoto2024}, which state that decoders based on the \emph{conditional} maximising and weighting distributions, respectively, are universal for (not necessarily additive) memoryless channels.
We note, however, that Theorem~\ref{thm:determistic-guessing-universality} is valid for more general additive unifilar channels (e.g. with additive Markov noise structure), and cannot be directly deduced from the aforementioned ones.

\subsubsection{Complexity}

In~\cite{duffy2019}, the asymptotic exponent of the number of queries needed in step 2 was identified using large deviation theory. Here, we proceed differently and obtain a non-asymptotic upper bound on this quantity using the method of types.

\begin{theorem}[Complexity] \label{thm:determinstic-guessing-complexity}
	(a)~For the maximising deterministic guessing decoder on additive unifilar channels, we have
	\begin{align} \label{eq:maximising-determinstic-guessing-complexity} 
		\frac{1}{n} \log Q_{\theta,G_m}(n) 
		\le \min\left\{ {H}_{1/2}(P_{\theta}) + 2\zeta(n) + \sigma_{\theta}(n),\ \log|\Acal|-R+\frac{1}{n} \right\}.
	\end{align}
	
	(b)~For the weighting deterministic guessing decoder on additive unifilar channels, we have 
	\begin{align} \label{eq:weighting-determinstic-guessing-complexity}
		\frac{1}{n} \log Q_{\theta,G_w}(n)
		\le \min\left\{ {H}_{1/2}(P_{\theta}) + \zeta(n) + \epsilon(n) + \sigma_{\theta}(n),\ \log|\Acal|-R+\frac{1}{n} \right\}.
	\end{align}
\end{theorem}

Theorem~\ref{thm:determinstic-guessing-complexity} partially recovers the result of \cite[Proposition~2]{duffy2019}, which gives the exact asymptotic exponent of $Q_{\theta,G_u}$ as $\min\{ {H}_{1/2}(P_{\theta}),\, \log|\Acal|-R \}$, when using GRAND with exact channel statistics, i.e. with $G_u = G^\star$. Our results, on the other hand, provide non-asymptotic upper bounds on that exponent, for guessing based on $G_m$ and $G_w$ under unknown channel law, both of which asymptotically match the exponent of~\cite[Proposition~2]{duffy2019}. By analysing the vanishing terms, we see that the guessing exponents behave as $\min\left\{ {H}_{1/2}(P_{\theta}) ,\ \log|\Acal|-R \right\} + O\left( (\log n)/n \right)$.

\subsection{Proofs}

\subsubsection{Universality}

Consider first decoders of the form~\eqref{eq:codeword-evaluation}. The average probability of error of such a decoder can be quantified in terms of the size of the equivocation set
\begin{equation*}
	\Ecal_u(\xv,\yv) \coloneqq \left\{ \xv' \in \Acal^n \suchthat u(\yv-\xv') \ge u(\yv-\xv) \right\}.
\end{equation*}
The next result shows how to compare the performance of two decoders in terms of these sets.

\begin{lemma}[{\hspace{0.001em}\cite{feder1998}}] \label{lemma:equivocation-sets}
	In the context of the notation above,
	\begin{equation}
		\frac{P_{\theta,{u}}(\error)}{P_{\theta,{\ML}}(\error)}
		\le \max_{(\xv,\yv) \in \Acal^n \times \Acal^n} \max \left\{ \frac{\left|\Ecal_{u}(\xv,\yv)\right|}{\left|\Ecal_{\ML}(\xv,\yv)\right|},\ 1 \right\}.
	\end{equation}
\end{lemma}
\begin{IEEEproof}
	The average probability of error can be written
	\begin{equation*}
		P_{\theta,{u}}(\error) = \sum_{\xv \in \Acal^n} \sum_{\yv \in \Acal^n} \frac{1}{|\Acal|^n} P_{\Zv}(\yv - \xv) \left[ 1 - \left( 1 - \frac{\left| \Ecal_{u}(\xv,\yv) \right|}{|\Acal|^n} \right)^{M-1} \right].
	\end{equation*}
	We thus have
	\begin{align*}
		\frac{P_{\theta,u}(\error)}{P_{\theta,\ML}(\error)}
		\le \max_{(\xv,\yv) \in \Acal^n \times \Acal^n} \frac{1 - \left( 1 - \frac{\left| \Ecal_{u}(\xv,\yv) \right|}{|\Acal|^n} \right)^{M-1}}{1 - \left( 1 - \frac{\left| \Ecal_{\ML}(\xv,\yv) \right|}{|\Acal|^n} \right)^{M-1}}
		\le \max_{(\xv,\yv) \in \Acal^n \times \Acal^n} \max \left\{ \frac{\left|\Ecal_{u}(\xv,\yv)\right|}{\left|\Ecal_{\ML}(\xv,\yv)\right|},\ 1 \right\},
	\end{align*}
	where the inequalities follow respectively from items 3 and 1 of~\cite[Lemma 2]{feder1998}.
\end{IEEEproof}

The following result provides a lower bound on the equivocation set under ML decoding.

\begin{lemma} \label{lemma:ml-decoder}
	For all $\xv,\yv \in \Acal^n$, we have
	\begin{equation}
		\left| \Ecal_{\ML}(\xv,\yv) \right|
		\ge 2^{n\left(H(\hat{P}_{\yv-\xv}) - \delta(n)\right)}.
	\end{equation}
\end{lemma}
\begin{IEEEproof}
	This result is the analogue of~\cite[Lemma~1]{ziv1985} in the context of additive channels, and the proof follows the same steps.	Using Lemma~\ref{lemma:method-of-types}, we have
	\begin{align*}
		\left| \Ecal_{\ML}(\xv,\yv) \right|
		&= \left| \left\{ \xv' \in \Acal^n \suchthat P_{\Zv}(\yv - \xv') \ge P_{\Zv}(\yv - \xv) \right\} \right|\\
		&\ge \left| \left\{ \xv' \in \Acal^n \suchthat P_{\Zv}(\yv - \xv') = P_{\Zv}(\yv - \xv) \right\} \right|\\
		&\ge \left| \left\{ \zv' \in \Acal^n \suchthat {\hat{P}_{\zv'} = \hat{P}_{\yv-\xv}} \right\} \right|\\
		&= \left| \Tcal_n(\hat{P}_{\yv-\xv}) \right|\\
		&\ge 2^{n\left({H}(\hat{P}_{\yv-\xv}) - \delta(n)\right)}.
	\end{align*}
\end{IEEEproof}

In decoding by deterministic noise guessing, when codeword $\xv$ is transmitted and $\yv$ is received, a decoding error occurs only if the codebook contains a codeword $\xv'$ such that $G_u(\yv - \xv') < G_u(\yv - \xv)$. 
Thus, the equivocation set for this decoder is~\cite{jouhed2024}
\begin{equation*}
	\Ecal_{G_u}(\xv,\yv) \coloneqq \left\{ \xv' \in \Acal^n \suchthat G_u(\yv - \xv') < G_u(\yv - \xv) \right\},
\end{equation*}
from which we immediately have
\begin{equation} \label{eq:bound-E-Gu}
	\left| \Ecal_{G_u}(\xv,\yv) \right| < G_{u}(\yv - \xv).
\end{equation}

\begin{lemma} \label{lemma:guessing-decoder}
	For all $\zv \in \Acal^n$, we have
	\begin{equation} \label{eq:guessing-maximising-decoder}
		G_m(\zv)
		\le 2^{n\left({H}(\hat{P}_{\zv}) + \zeta(n) \right)}.
	\end{equation}
	and
	\begin{equation} \label{eq:guessing-weighting-decoder}
		G_w(\zv)
		\le 2^{n\left({H}(\hat{P}_{\zv}) +  \epsilon(n) \right)}.
	\end{equation}
\end{lemma}
\begin{IEEEproof}
	For the first part, we have
	\begin{align*}
		G_m(\zv)
		&\le \left|\left\{ \zv' \in \Acal^n \suchthat P_m(\zv') \ge P_m(\zv) \right\}\right|\\
		&= \left|\left\{ \zv' \in \Acal^n \suchthat {H}(\hat{P}_{\zv'}) \le {H}(\hat{P}_{\zv}) \right\}\right|\\
		&= \sum_{\hat{P}_Z \in \Pcal_n(\Acal) \suchthat {H}(\hat{P}_Z) \le {H}(\hat{P}_{\zv})} \left| \Tcal_n(\hat{P}_{Z}) \right|\\
		&\le 2^{n \left(\zeta(n) + {H}(\hat{P}_{\zv})\right)},
	\end{align*}
	where the last inequality follows from Lemma~\ref{lemma:method-of-types}.
	
	For the second part, we have
	\begin{align*}
		G_w(\zv)
		&\le \left|\left\{ \zv' \in \Acal^n \suchthat P_w(\zv') \ge P_w(\zv) \right\}\right|\\
		&\le \frac{1}{P_w(\zv)}\\
		&\le \frac{2^{n\epsilon(n)}}{\hat{P}_{\zv}(\zv)}\\
		&= 2^{n\left( {H}(\hat{P}_{\zv}) + \epsilon(n) \right)},
	\end{align*}
	and the last inequality follows from Lemma~\ref{lemma:weighting}.
\end{IEEEproof}

\begin{IEEEproof}[Proof of Theorem~\ref{thm:determistic-guessing-universality}]
	Combining the results of Lemmas~\ref{lemma:equivocation-sets}, \ref{lemma:ml-decoder}, \ref{lemma:guessing-decoder}, and \eqref{eq:bound-E-Gu}, we have
	\begin{align*}
		\frac{1}{n}\log\frac{P_{\theta,G_m}(\error)}{P_{\theta,\ML}(\error)}
		&\le \frac{1}{n} \log \left( \max_{(\xv,\yv)\in\Acal^n\times\Acal^n} \frac{\left| \Ecal_{G_m}(\xv,\yv) \right|}{\left| \Ecal_{\ML}(\xv,\yv) \right|} + 1 \right)\\
		&\le \frac{1}{n} \log \left( 2^{n\left( \zeta(n) + \delta(n) \right)} + 1 \right)\\
		&\le \frac{1}{n} \log \left( 2 \cdot 2^{n\left(\zeta(n)+\delta(n)\right)} \right)\\
		&= \frac{1}{n} + \zeta(n) + \delta(n),
	\end{align*}
	which goes to $0$ as $n\to\infty$ and does not depend on $\theta \in \Theta$. An analogous result holds for $P_{\theta,G_w}$, using the second part of Lemma~\ref{lemma:guessing-decoder}.
\end{IEEEproof}

\subsubsection{Complexity}

Since the messages are equiprobable, it is without loss of generality to consider that message $m=1$ was sent. The algorithm stops when either the correct codeword $\xv_1$, or an incorrect one $\xv_2,\dots,\xv_M$ is found. The average number of such queries (over random codes and noise realisations) when using guessing based on the ranking function $G_u$ in channel $\theta$ is then
\begin{align*}
	Q_{\theta,G_u}(n)
	&= \E_{\Xv_1,\dots,\Xv_M,\Yv} \left[ \min\left\{ G_u(\Yv-\Xv_1),\ \min_{m'\neq1} G_u(\Yv - \Xv_{m'}) \right\} \right]\\
	&\le \min\left\{
	\E_{\Xv_1,\Yv} \left[  G_u(\Yv-\Xv_1)  \right],\, \E_{\Xv_2,\dots,\Xv_M,\Yv} \left[ \min_{m'\neq1} G_u(\Yv - \Xv_{m'}) \right]
	\right\}.
\end{align*}

Since $\yv = \xv_1 + \zv$, the first term inside the minimum is $\E_{\Xv_1,\Yv} \left[  G_u(\Yv-\Xv_1)  \right] = \E_{\Zv} \left[G_u(\Zv)\right]$. For a given $\yv$, due to random coding, each $\yv - \xv_{m'}$ is uniformly distributed in $\Acal^n$. Thus $K_{m'} \coloneqq G(\yv-\xv_{m'})$ is uniformly distributed in $\{ 1, \dots, |\Acal|^n \}$ and independent of $\yv$. Denoting $K \coloneqq \min_{m' \neq 1} K_{m'}$, the second term becomes $\E_{\Xv_2,\dots,\Xv_M,\Yv} \left[ \min_{m'\neq1} G_u(\Yv - \Xv_{m'}) \right] = \E_K \left[K\right]$. Combining these results yields
\begin{align} \label{eq:complexity-deterministic-guessing}
	Q_{\theta,G_u}(n)
	\le \min\left\{ E_{\Zv} \left[G_u(\Zv)\right],\, \E_K \left[ K \right] \right\}.
\end{align}

\begin{lemma} \label{lemma:expectation-B}
	Let $K_{m'} \sim \Ucal \left( \{ 1, \dots, |\Acal|^n \} \right)$, for $m' \in \left\{2, \dots, M\right\}$, $M = 2^{nR}$, and $K \coloneqq \min_{m' \neq 1} K_{m'}$. Then,
	\begin{align}
		2^{n\left( \log|\Acal| - R \right)}
		\le \E_K\left[K\right]
		< 2^{n\left(\log|\Acal|-R + \frac{1}{n}\right)}.
	\end{align}
\end{lemma}
\begin{IEEEproof}
	See Appendix~\ref{sec:appendix-proof-lemma-expectation}.
\end{IEEEproof}

\begin{lemma} \label{lemma:expectation-2nH}
	We have
	\begin{equation} \label{eq:expectation-exp-n-entropy}
		\E_{\Zv}\left[ 2^{n {H}(\hat{P}_{\Zv})} \right] \le 2^{n\left( {H}_{1/2}(P_{\theta}) + \zeta(n) + \sigma_{\theta}(n) \right)}.
	\end{equation}
\end{lemma}
\begin{IEEEproof}
	We have
	\begin{align}
		\E_{\Zv}\left[ 2^{n{H}(\hat{P}_{\Zv})} \right]
		&= \sum_{\zv\in\Acal^n} P_{\Zv}(\zv) 2^{n{H}(\hat{P}_{\zv})} \nonumber\\
		&= \sum_{\hat{P}_{Z} \in \Pcal_n(\Acal)} \left| \Tcal_n(\hat{P}_{Z}) \right| 2^{-n\left( {H}(\hat{P}_{Z}) + {D}(\hat{P}_{Z} \| P_{\Zv}) \right)} 2^{n{H}(\hat{P}_{Z})} \nonumber\\
		&\le \sum_{\hat{P}_{Z} \in \Pcal_n(\Acal)} 2^{n\left( {H}(\hat{P}_{Z}) - {D}(\hat{P}_{Z} \| P_{\Zv} ) \right)} \nonumber\\
		&\le 2^{n\zeta(n)} 2^{n \max_{\hat{P}_{Z} \in \Pcal_n(\Acal)} \left( {H}(\hat{P}_{Z}) - {D}(\hat{P}_{Z} \| P_{\Zv} ) \right)} \nonumber\\
		&\le 2^{n\left( {H}_{1/2}(P_{\theta}) + \zeta(n) + \sigma_{\theta}(n) \right)} \label{eq:expectation-exp-n-entropy},
	\end{align}
	where the first two inequalities come from Lemma~\ref{lemma:method-of-types}, and the third one, from Lemma~\ref{lemma:bound-variational-characterisation}, with $\alpha=1/2$.
\end{IEEEproof}

\begin{IEEEproof}[Proof of Theorem~\ref{thm:determinstic-guessing-complexity}]
	Combining the results of Lemmas~\ref{lemma:guessing-decoder} and \ref{lemma:expectation-2nH}, we have
	\begin{equation} \label{eq:maximising-deterministic-guessing-complexity}
		\E_{\Zv} \left[G_m(\Zv)\right] \le 2^{n \left( {H}_{1/2}(P_{\theta}) + 2 \zeta(n) + \sigma_{\theta}(n) \right)}.
	\end{equation}
	and
	\begin{equation} \label{eq:weighting-deterministic-guessing-complexity}
		\E_{\Zv}\left[G_w(\Zv)\right] \le 2^{n \left( {H}_{1/2}(P_{\theta}) + \zeta(n) + \epsilon(n) + \sigma_{\theta}(n) \right)}.
	\end{equation}
	Those, combined with Lemma~\ref{lemma:expectation-B} and applied to~\eqref{eq:complexity-deterministic-guessing} yield the desired result.
\end{IEEEproof}

\section{Decoding by Randomised Noise Guessing} \label{sec:randomised-guessing}

While deterministic noise guessing can significantly reduce the average decoding complexity, it still requires enumerating all types, whose number grows exponentially with the number of states of the channel process. This motivates us to consider an even simpler strategy based on randomised noise guessing~\cite{merhav2020,hanawal2010,boztas2012,huleihel2017}.
Instead of generating noise guesses in a fixed (deterministic) order, the guesses are drawn according to some distribution~$\tilde{P}_Z$ (randomised). Specifically, given $\yv$, one repeatedly draws samples $\zv \sim \tilde{P}_Z$ and queries whether $\yv - {\zv} \in \Ccal$ until a codeword is found.
Decoding by randomised noise guessing can be seen as the guessing decoding analogue of the stochastic decoder~\cite{scarlett2015,merhav2017}.
We will study \emph{maximising randomised guessing decoding} based on the maximising distribution $\tilde{P}_m(\zv) \coloneqq P_m(\zv)$, and \emph{weighting randomised guessing decoding} based on the weighting distribution $\tilde{P}_{w}(\zv) \coloneqq {P}_{w}(\zv)$.

\subsection{Decoding Scheme} \label{subsec:randomised-guessing-scheme}

Similarly to the decoding by deterministic noise guessing described in Section~\ref{subsec:deterministic-guessing-scheme}, here too there are two main steps involved in the proposed decoding scheme. 

\paragraph*{Step~1: Sample the guesses}

Different strategies are possible to sample noise sequences; the first two of them are based on~\cite{merhav2020}.
\begin{enumerate}[label=(\roman*)]
	\item If $\tilde{P}_{Z}(\zv)$ only depends on $\zv$ via its type $\hat{P}_{\zv}$, sampling can be done in two steps: first draw a type, according to $\tilde{P}_Z\big(\Tcal_n(\hat{P}_{\zv})\big)$ (to be computed in advance), and then uniformly draw a sequence from that type. Computing the $\tilde{P}_Z$-probabilities of all types requires $O\big(n^{|\Acal||\Scal|}\big)$ computations and $O\big(n^{|\Acal||\Scal|}\big)$ storage. Drawing a sequence from the type class $\Tcal_n(\hat{P}_{\zv})$ can be done by drawing an index from $1$ to $|\Tcal_n(\hat{P}_Z)|$, and then using an enumeration algorithm (cf. Section~\ref{subsec:deterministic-guessing-scheme}) to get the corresponding sequence.
	
	\item The structure of the weighting distribution $P_w = P_{\KT}$ with Dirichlet priors allows sequential sampling of the noise sequences~$\zv$. Specifically,
		\begin{equation}
			P_\KT\big( z_i \given z_{1}^{i-1} \big)
			= \frac{a_{z_1^{i-1}}(z_i,s_{i-1}) + \frac{1}{2}}{a_{s_1^{i-1}}(s_{i-1}) + \frac{|\Acal|}{2}},
		\end{equation}
		where $s_i = f(z_i,s_{i-1})$. Supposing that the function $f$ is implemented in a table, this requires $O(|\Acal|n)$ computations and $O(|\Acal||\Scal|\log n)$ storage.
	
	\item As $P_w$ is a mixture distribution, a two-step procedure can be applied: first draw a parameter $\theta\sim\pi$ according to the prior distribution, and then draw a sequence $\zv \sim P_{\theta}$. The complexity of this is $O(|\Acal||\Scal|n)$ computations and $O(|\Acal||\Scal|)$ storage. Note that this can be done even if the prior $\pi$ does not yield a closed-form expression for the weighting distribution $P_w$.
\end{enumerate}

\paragraph*{Step~2: Query the sequences}

Same as step~2 from Section~\ref{subsec:deterministic-guessing-scheme}.

\subsection{Main Results}

We denote $P_{\theta,\tilde{P}_Z}(\error)$ the average probability of error and $Q_{\theta,\tilde{P}_Z}(\error)$ the average number of guesses needed to find a codeword when decoding with randomised guessing with guess distribution $\tilde{P}_Z$. In the following, the main results are stated, their proofs being postponed to the end of the section. 

\subsubsection{Universality}

Codeword $\xv_m$ will be selected whenever it is the first valid one encountered by randomly drawing noise sequences~$\zv$ and subtracting those from $\yv$. Say this happens at the $k$-th trial: this means that all previous $k-1$ noise sequences that were tested do not correspond to any codeword at all. We can thus write the probability of selecting message $m$ as
\begin{align} \label{eq:randomised-decoding-distribution}
	\Pbb\left(\hat{m} = m \given \yv\right)
	&= \sum_{k=1}^{\infty} {\left( 1- \sum_{m'=1}^{M} \tilde{P}_Z(\yv - \xv_{m'}) \right)}^{k-1} \tilde{P}_{Z}(\yv - \xv_m) \nonumber\\
	&= \frac{\tilde{P}_{Z}(\yv - \xv_m)}{\sum_{m'=1}^{M} \tilde{P}_Z(\yv - \xv_{m'})},
\end{align}
where the geometric sum converges because $\sum_{m'=1}^{M} \tilde{P}_Z(\yv-\xv_{m'}) > 0$.

Since the messages are equiprobable, we can again assume that $m=1$, without loss of generality. If $\xv_1$ was transmitted and $\yv$ received, correct decoding occurs if and only if $\hat{\xv} = \xv_1$ is chosen. The probability of error is the probability of the complementary event. Taking the averages (over random codes and noise realisations), we get
\begin{equation} \label{eq:random-guessing-error-probability}
	P_{\theta,\tilde{P}_Z}(\error)
	= \E_{\Xv_1,\dots,\Xv_M,\Yv} \left[ 1 -  \frac{\tilde{P}_{Z}(\Yv - \Xv_1)}{\sum_{m'=1}^{M} \tilde{P}_Z(\Yv - \Xv_{m'})} \right].
\end{equation}

\begin{theorem}[Universality] \label{thm:randomised-guessing-universality}
	(a)~The maximising randomised guessing decoder is random-coding universal on additive unifilar channels, with
	\begin{equation} \label{eq:randomised-maximising-guessing-universality}
		\frac{1}{n}\log\frac{P_{\theta,\tilde{P}_m}(\error)}{P_{\theta,\ML}(\error)}
		\le \frac{2}{n} + \zeta(n) + \delta(n).
	\end{equation}
	
	(b)~The weighting randomised guessing decoder  is random-coding universal on additive unifilar channels, with
	\begin{equation} \label{eq:randomised-weighting-guessing-universality}
		\frac{1}{n}\log\frac{P_{\theta,\tilde{P}_w}(\error)}{P_{\theta,\ML}(\error)}
		\le \frac{2}{n} + \epsilon(n) + \delta(n). 
	\end{equation}
\end{theorem}

Note that~\eqref{eq:random-guessing-error-probability} coincides with the probability of error of a stochastic decoder~\cite{scarlett2015,merhav2017} that chooses codewords according to~\eqref{eq:randomised-decoding-distribution}. In this sense, the first part of Theorem~\ref{thm:randomised-guessing-universality}, specialised to memoryless additive channels, is somewhat analogous to the result of~\cite[p.~5043]{merhav2017} stating that on (not necessarily additive) memoryless channels, a stochastic decoder (equivalent to one) based on a \emph{conditional} maximising distribution is universal. Moreover, note that the results of Theorem~\ref{thm:randomised-guessing-universality} are nearly the same as those of Theorem~\ref{thm:determistic-guessing-universality}, except by replacing $1/n$ by $2/n$; in particular, the redundancies scale with $O\left( (\log n)/n \right)$ here too. Consequently, there is no loss in universality by using randomised guessing instead of deterministic guessing based on maximising and weighting distributions.

\subsubsection{Complexity}

We now proceed to bound the complexity of randomised guessing decoding.

\begin{theorem}[Complexity] \label{thm:randomised-guessing-complexity}	
	(a)~For the maximising randomised guessing decoder on additive unifilar channels, there exists an $n_0 \coloneqq n_0(|\Acal|,R)$ such that, for $n \ge n_0$,
	\begin{align} \label{eq:weihting-randomised-guessing-complexity}
		&\frac{1}{n} \log {Q}_{\theta,\tilde{P}_m}(n)
		\le
		\min\vast\{
			{H}_{1/2}(P_\theta) + 2 \zeta(n) + \sigma_{\theta}(n),\ \nonumber\\
			&\hspace{10.5em}\max\left\{
			\lambda_1(n),\ 
			\log|\Acal|-R+ \lambda_2(n)
			\right\} + \zeta(n) + \frac{1}{n}
		\vast\}.
	\end{align} 
	
	(b)~For the weighting randomised guessing decoder on additive unifilar channels, there exists an $n_0 \coloneqq n_0(|\Acal|,R)$ such that, for $n \ge n_0$,
	\begin{align} \label{eq:weihting-randomised-guessing-complexity}
		&\frac{1}{n} \log {Q}_{\theta,\tilde{P}_w}(n)
		\le
		\min\vast\{
			{H}_{1/2}(P_\theta) + \zeta(n) + \epsilon(n) + \sigma_{\theta}(n),\ \nonumber\\
			&\hspace{10.5em}\max\left\{
			\lambda_1(n),\ 
			\log|\Acal|-R+\lambda_2(n)
			\right\} + \epsilon(n) + \frac{1}{n}
		\vast\}.
	\end{align}

	In these equations, $\lambda_1(n) = O\left( {(\log n)}/{n} \right)$ and $\lambda_2(n) = O\left( {1}/{n^2} \right)$; exact expressions for both quantities appear in the proof.
\end{theorem}

Similarly to Theorem~\ref{thm:determinstic-guessing-complexity}, Theorem~\ref{thm:randomised-guessing-complexity} provides non-asymptotic upper bounds on the complexity exponent of randomised guessing strategies. Comparing them, we see that, for large enough $n$, the upper bounds are essentially the same when decoding with either deterministic or randomised guessing, a result that mimics the fact that universal randomised guessing is possible~\cite{merhav2020}. As far as complexity exponents are concerned, there is no essential loss in adopting randomised guessing decoding as compared to the deterministic guessing counterpart. In particular, here too the asymptotic behaviour is $\min\left\{ {H}_{1/2}(P_{\theta}) ,\ \log|\Acal|-R \right\} + O\left( (\log n)/n \right)$.

We conclude this subsection with a comparison between decoding by deterministic and randomised guessing. As we have seen, the redundancy and complexity exponents of both strategies are asymptotically the same.
While randomised noise guessing typically requires more queries than deterministic guessing due to its random nature, it can present advantages for practical implementation.
First, drawing sequences with strategies (ii) or (iii) for the step~1 of randomised guessing decoding (cf. Section~\ref{subsec:randomised-guessing-scheme}) has linear complexity in both the block-length~$n$ and the number of states~$|\Scal|$. This is in contrast to the step~1 of deterministic guessing decoding (cf. Section~\ref{subsec:deterministic-guessing-scheme}) or strategy (i) for step 1 of randomised guessing decoding, which require enumerating the types. The cost of this operation scales with $O\big( n^{|\Acal||\Scal|} \big)$, which, although polynomial in $n$, grows exponentially with $|\Scal|$, rendering it impractical with a large number of states.
Moreover, the randomised strategy is highly parallelisable, as the module that produces the guesses can be replicated without any need for coordination between them, as in~\cite{merhav2020}.
Finally, we note that strategies (ii) and (iii) for the step~1 of randomised guessing decoding only work for the weighting distribution, showing an advantage of that over the maximising distribution.

\subsection{Proofs}

\subsubsection{Universality}

The next lemma provides a sufficient condition to compare the performance of a randomised guessing decoder with the optimal ML decoder.

\begin{lemma} \label{lemma:universality-randomised-guessing}
	Let $\Delta>0$. If
	\begin{equation} \label{eq:hypothesis-lemma-universality-randomised-guessing}
		\frac{\E_{\bar{\Xv}}\left[\tilde{P}_{Z}(\yv-\bar{\Xv})\right]}{\tilde{P}_{Z}(\yv-\xv)}
		\le \frac{\left| \Ecal_{\ML}(\xv,\yv) \right|}{|\Acal|^n} \cdot  2^{n\Delta},
	\end{equation}
	for all $(\xv,\yv) \in \Acal^n \times \Acal^n$, then
	\begin{equation}
		\frac{P_{\theta,\tilde{P}_Z}(\error)}{P_{\theta,\ML}(\error)} \le 4 \cdot 2^{n\Delta},
	\end{equation}
	where the expectation is with respect to the distribution $\bar{\Xv} \sim P_{\Xv}(\bar\xv) = \frac{1}{|\Acal|^n}$.
\end{lemma}

\begin{IEEEproof}
	Since the randomised guessing decoder is equivalent to a stochastic decoder that chooses messages according to~\eqref{eq:randomised-decoding-distribution} (compare with \cite[Equation~(1)]{scarlett2015}), we can use \cite[Theorem~1]{scarlett2015} with $s=1$ to get
	\begin{equation*} 
		P_{\theta,\tilde{P}_Z}(\error) \le
		\E_{\Xv,\Yv}\left[
		\min\left\{ 1,\
		(M-1)\frac{
			\E_{\bar{\Xv}}\left[\tilde{P}_{Z}(\Yv-\bar{\Xv}) \mgiven \Yv \right]
		}{
			\tilde{P}_Z(\Yv-\Xv)
		} \right\}
		\right].
	\end{equation*}
	
	On the other hand, for the ML decoder, we can apply the lower bound on the RCU bound~\cite[Theorem~1]{scarlett2014} with
	\begin{equation*}
		\Pbb_{\bar{\Xv}}\left[ P_{\Zv}(\yv-\bar{\Xv}) \ge P_{\Zv}(\yv-\xv) \right]
		= \frac{|\Ecal_{\ML}(\xv,\yv)|}{|\Acal|^n},
	\end{equation*}
	which holds for uniform random coding over $\Acal^n$, to obtain
	\begin{equation*}
		P_{\theta,\ML}(\error)
		\ge \frac{1}{4}\E_{\Xv,\Yv}\left[\min\left\{1,\ (M-1)\frac{\left|\Ecal_\ML(\Xv,\Yv)\right|}{\left|\Acal\right|^n}\right\}\right].
	\end{equation*}
	
	Combining these inequalities with the hypothesis \eqref{eq:hypothesis-lemma-universality-randomised-guessing} yields the desired result.
\end{IEEEproof}

\begin{IEEEproof}[Proof of Theorem~\ref{thm:randomised-guessing-universality}]
	In view of Lemma~\ref{lemma:universality-randomised-guessing}, we only need to show that
	\begin{equation} \label{eq:randomised-maximising-guessing}
		\frac{\E_{\bar{\Xv}}\left[\tilde{P}_{m}(\yv-\bar{\Xv})\right]}{\tilde{P}_m(\yv-\xv)}
		\le \frac{\left| \Ecal_{\ML}(\xv,\yv) \right|}{|\Acal|^n} \cdot 2^{n\left( \zeta(n) + \delta(n) \right)}.
	\end{equation}
	and
	\begin{equation} \label{eq:randomised-weighting-guessing}
		\frac{\E_{\bar{\Xv}}\left[\tilde{P}_{w}(\yv-\bar{\Xv})\right]}{\tilde{P}_{w}(\yv-\xv)}
		\le \frac{\left| \Ecal_{\ML}(\xv,\yv) \right|}{|\Acal|^n} \cdot  2^{n\left( \epsilon(n) + \delta(n) \right)}.
	\end{equation}

	Indeed, for~\eqref{eq:randomised-maximising-guessing}, we have
	\begin{align*}
		\frac{\E_{\bar{\Xv}}\left[\tilde{P}_{m}(\yv-\bar{\Xv})\right]}{\tilde{P}_m(\yv-\xv)}
		&= \frac{\sum_{\bar{\xv}\in\Acal^n} \frac{1}{\left|\Acal\right|^n} 2^{-n{H}(\hat{P}_{\yv-\bar{\xv}})} }{2^{-n{H}(\hat{P}_{\yv-\xv})}}\\
		&= \frac{1}{|\Acal|^n} \frac{\sum_{\hat{P}_Z \in \Pcal_n(\Acal)} \left| \Tcal_n(\hat{P}_Z) \right| 2^{-n{H}(\hat{P}_Z)} }{2^{-n{H}(\hat{P}_{\yv-\xv}) }}\\
		&\le \frac{1}{|\Acal|^n} \left| \Pcal_n(\Acal) \right| 2^{n{H}(\hat{P}_{\yv-\xv})}\\
		&\le \frac{\left| \Ecal_{\ML}(\xv,\yv) \right|}{|\Acal|^n} 2^{n\left( \zeta(n)+\delta(n) \right)},
	\end{align*}
	where in the first inequality we used Lemma~\ref{lemma:method-of-types}, and in the second one we used Lemmas~\ref{lemma:method-of-types} and \ref{lemma:ml-decoder}.
	
	Meanwhile, for~\eqref{eq:randomised-weighting-guessing}, we have
	\begin{align*}
		\frac{\E_{\bar{\Xv}}\left[\tilde{P}_{w}(\yv-\bar{\Xv})\right]}{\tilde{P}_{w}(\yv-\xv)}
		&= \frac{\sum_{\bar{\xv}\in\Acal^n} \frac{1}{\left|\Acal^n\right|} P_{w}(\yv-\bar{\xv}) }{P_{w}(\yv-\xv)}\\
		&= \frac{1}{|\Acal|^n} \frac{1}{P_{w}(\yv-\xv)}\\
		&\le \frac{1}{|\Acal|^n} 2^{n\epsilon(n)} 2^{n{H}(\hat{P}_{\yv-\xv}) }\\
		&\le \frac{\left| \Ecal_{\ML}(\xv,\yv) \right|}{|\Acal|^n} 2^{n\left( \epsilon(n)+\delta(n) \right)},
	\end{align*}
	where in the first inequality we used Lemma~\ref{lemma:weighting}, and in the second we used Lemma~\ref{lemma:ml-decoder} again.
\end{IEEEproof}

\subsubsection{Complexity}

For a code $\Ccal = \left( \xv_1, \dots, \xv_M \right)$, the average number of guesses needed to find a codeword follows a geometric distribution with success probability $\tilde{P}_{Z}\left( \bigcup_{m'=1}^{M} \left\{\yv - \xv_{m'}\right\} \right)$; the average number of queries (with respect to the randomised guessing) is thus $\left[ \tilde{P}_{Z}\left( \bigcup_{m'=1}^{M} \left\{\yv - \xv_{m'}\right\} \right) \right]^{-1}$. Note that, in general, the probability of the union is not the sum of probabilities of each sequence, unless all codewords are distinct. Let us denote $K_M \coloneqq K_M(\Ccal) \coloneqq \left| \bigcup_{m'=1}^{M} \left\{ \xv_{m'} \right\} \right|$ the number of distinct codewords in the code. If $K_M = k$, then we can extract a sequence $\left( \xv_1^*, \dots, \xv_k^* \right) \subset \Ccal$ of distinct codewords, with $\xv_1^* = \xv_1$. Again assuming that message $m=1$ was sent, we can write the average number of queries (with respect to random codes and noise realisations) as
\begin{align} 
	Q_{\theta,\tilde{P}_Z}(n)
	&= \E_{\Xv_1,\dots,\Xv_M,\Yv} \left[
	\frac{1}{\tilde{P}_{Z}\left( \bigcup_{m'=1}^{M} \left\{\Yv - \Xv_{m'}\right\} \right)}
	\right] \nonumber\\
	&= \sum_{k=1}^{M} \Pbb\left( K_M = k \right) \E_{\Xv_1,\dots,\Xv_M,\Yv} \left[
	\frac{1}{\sum_{m'=1}^{k} \tilde{P}_{Z}\left( \Yv - \Xv^{*}_{m'} \right)}
	\mgiven \Xv_1^* \neq \cdots \neq \Xv_k^*
	\right]. \label{eq:complexity-randomised-guessing-minimum-expression}
\end{align}

\begin{lemma} \label{lemma:randomised-guessing-complexity-1}
	We have
	\begin{equation} \label{eq:maximising-randomised-guessing-complexity-1}
		\E_{\Zv} \left[ \frac{1}{\tilde{P}_{m}(\Zv)} \right]
		\le 2^{n\left( {H}_{1/2}(P_{\theta}) + 2\zeta(n) + \sigma_{\theta}(n) \right)}
	\end{equation}
	and
	\begin{equation} \label{eq:weighting-randomised-guessing-complexity-1}
		\E_{\Zv} \left[ \frac{1}{\tilde{P}_{w}(\Zv)} \right]
		\le 2^{n\left( {H}_{1/2}(P_{\theta}) + \zeta(n) + \epsilon(n) + \sigma_{\theta}(n) \right)}.
	\end{equation}
\end{lemma}
\begin{IEEEproof}
	For the first part, we have
	\begin{align}
		\E_{\Zv} \left[ \frac{1}{\tilde{P}_{m}(\Zv)} \right]
		= \E_{\Zv} \left[ \frac{\sum_{\bar{\zv}\in\Acal^n} 2^{-n{H}(\hat{P}_{\bar{\zv}})} }{2^{-n{H}(\hat{P}_{\Zv})}} \right]
		= \left( \sum_{\bar{\zv}\in\Acal^n} 2^{-n{H}(\hat{P}_{\bar{\zv}})}  \right) \E_{\Zv} \left[ 2^{n{H}(\hat{P}_{\Zv})} \right]. \label{eq:randomised-guessing-complexity-1-initial}
	\end{align}
	For the term in parenthesis, we have, as in~\cite[p.~118]{merhav2020},
	\begin{align}
		\sum_{\bar{\zv} \in \Acal^n} 2^{-n{H}(\hat{P}_{\bar{\zv}})}
		&= \sum_{\bar{\zv} \in \Acal^n} \max_{\theta\in\Theta} P_{\Zv,\theta}(\bar{\zv}) \nonumber\\
		&= \sum_{\bar{\zv} \in \Acal^n} \max_{\hat{P}_Z \in \Pcal_n(\Acal)} \hat{P}_{Z}(\bar{\zv})\nonumber\\
		&\le \sum_{\bar{\zv} \in \Acal^n} \sum_{\hat{P}_Z \in \Pcal_n(\Acal)} \hat{P}_Z(\bar{\zv}) \nonumber\\
		&= \left|\Pcal_n(\Acal)\right| \nonumber\\
		&\le  2^{n\zeta(n)} \label{eq:sum-exp-n-entropy},
	\end{align}
	thanks to Lemma~\ref{lemma:method-of-types}. Applying \eqref{eq:sum-exp-n-entropy} and Lemma~\ref{lemma:expectation-2nH} to \eqref{eq:randomised-guessing-complexity-1-initial}, we get the desired result.

	For the second part, analogous steps with Lemma~\ref{lemma:weighting} yield
	\begin{align*}
		\E_{\Zv} \left[ \frac{1}{\tilde{P}_{w}(\Zv)} \right]
		\le 2^{n \epsilon(n) } \E_{\Zv} \left[ 2^{n{H}(\hat{P}_{\Zv})} \right]
		\le 2^{n\left( {H}_{1/2}(P_{\theta}) + \zeta(n) + \epsilon(n) + \sigma_{\theta}(n) \right)}.
	\end{align*}
\end{IEEEproof}

\begin{lemma} \label{lemma:randomised-guessing-complexity-2}
	For any distribution on $(\Xv_1, \dots, \Xv_k) \in \left(\Acal^n\right)^k$, we have
	\begin{equation} \label{eq:maximising-randomised-guessing-complexity-2}
		\E_{\Xv_1,\dots,\Xv_k} \left[
		\frac{1}{\sum_{i=1}^{k} \tilde{P}_{m}(\Xv_i)} \mgiven \Xv_1 \neq \cdots \neq \Xv_k
		\right]
		\le  2^{n\zeta(n)} \frac{|\Acal|^n}{k},
	\end{equation}
	and
	\begin{equation} \label{eq:weighting-randomised-guessing-complexity-2}
		\E_{\Xv_1,\dots,\Xv_k} \left[
		\frac{1}{\sum_{i=1}^{k} \tilde{P}_{w}(\Xv_i)} \mgiven \Xv_1 \neq \cdots \neq \Xv_k
		\right]
		\le  2^{n \epsilon(n) } \frac{|\Acal|^n}{k}.
	\end{equation}
\end{lemma}
\begin{IEEEproof}
	For the first part, using Lemma~\ref{lemma:method-of-types} and \eqref{eq:sum-exp-n-entropy}, we have
	\begin{align*}
		&\hspace{-2em}\E_{\Xv_1,\dots,\Xv_k}\left[
			\frac{1}{\sum_{i=1}^{k} \tilde{P}_{m}(\Xv_i)}
			\mgiven \Xv_1 \neq \cdots \neq \Xv_k
		\right]\\
		&=
		\left( \sum_{\bar{\xv}\in\Acal^n} 2^{-nH(\hat{P}_{\bar{\xv}})} \right)
		\E_{\Xv_1,\dots,\Xv_k}\left[
			\frac{1}{\sum_{i=1}^{k} 2^{-n H(\hat{P}_{\Xv_i}) }}
			\mgiven \Xv_1 \neq \cdots \neq \Xv_k
		\right]\nonumber\\
		&\le 2^{n\zeta(n)} \E_{\Xv_1,\dots,\Xv_k}\left[
			\frac{1}{\sum_{i=1}^{k} 2^{-n \log|\Acal| }}
			\mgiven \Xv_1 \neq \cdots \neq \Xv_k
		\right]\\
		&= 2^{n\zeta(n)} \frac{|\Acal|^n}{k}.
	\end{align*}

	The second part is analogous, with an application of Lemma~\ref{lemma:weighting}.
\end{IEEEproof}

\begin{lemma} \label{lemma:bound-probability-Km}
	For a code $\Ccal = \left( \xv_1, \dots, \xv_M \right)$, let $K_M \coloneqq K_M(\Ccal) \coloneqq \left| \bigcup_{m'=1}^{M} \left\{ \xv_{m'} \right\} \right|$ and $M = 2^{nR}$. If $1 \le k^* < \E\left[K_M\right]$, then
	\begin{equation}
		\Pbb\left( K_M \le k^{*} \right)
		\le \frac{1}{|\Acal|^n} \left( \frac{3 \cdot 2^{nR}}{\E\left[ K_M \right]-k^*} \right)^2.
	\end{equation}
\end{lemma}
\begin{IEEEproof}
	See Appendix~\ref{sec:appendix-proof-lemma-probability}.
\end{IEEEproof}

\begin{IEEEproof}[Proof of Theorem~\ref{thm:randomised-guessing-complexity}]
	The proof consists in showing two upper bounds for the complexities $Q_{\theta,\tilde{P}_m}(n)$ and $Q_{\theta,\tilde{P}_w}(n)$. We will treat part~(a), part~(b) being analogous. First, noting that $\tilde{P}_{Z}\left( \bigcup_{m'=1}^{M} \left\{ \yv - \xv_{m'} \right\} \right) \ge \tilde{P}_{Z}\left( \yv - \xv_{1} \right)$, we can bound \eqref{eq:complexity-randomised-guessing-minimum-expression} by
	\begin{align*} 
		Q_{\theta,\tilde{P}_Z}(n)
		\le \E_{\Xv_1,\Yv} \left[
			\frac{1}{\tilde{P}_{Z}\left( \Yv - \Xv_{1} \right)}
		\right]
		= \E_{\Zv} \left[
			\frac{1}{\tilde{P}_{Z}\left( \Zv \right)}
		\right].
	\end{align*}
	Then, thanks to Lemma~\ref{lemma:randomised-guessing-complexity-1}, we get, for the maximising distribution,
	\begin{align} 
		Q_{\theta,\tilde{P}_m}(n)
		\le \E_{\Zv} \left[
			\frac{1}{\tilde{P}_{m}\left( \Zv \right)}
		\right]
		\le 2^{n\left( {H}_{1/2}(P_\theta) + 2\zeta(n) + \sigma_{\theta}(n) \right)}. \label{eq:proof-first-upper-bound}
	\end{align}
	
	For the second bound, apply Lemma~\ref{lemma:randomised-guessing-complexity-2} to~\eqref{eq:complexity-randomised-guessing-minimum-expression} to get
	\begin{align*} 
		Q_{\theta,\tilde{P}_m}(n)
		&= \sum_{k=1}^{M} \Pbb\left( K_M = k \right)
			\E_{\Xv_1,\dots,\Xv_M,\Yv} \left[
			\frac{1}{\sum_{m'=1}^{k} \tilde{P}_{m}\left( \Yv - \Xv^{*}_{m'} \right)}
			\mgiven \Xv_1^* \neq \cdots \neq \Xv_k^*
		\right]\\
		&\le 2^{n \zeta(n) } \sum_{k=1}^{M} \Pbb\left( K_M = k \right)
		\frac{|\Acal|^n}{k}.
	\end{align*}	
	We can split the sum and bound it as
	\begin{align}
		\sum_{k=1}^{M} \Pbb\left( K_M=k \right) \frac{|\Acal|^n}{k}
		&= \sum_{k=1}^{k^*} \Pbb\left( K_M=k \right) \frac{|\Acal|^n}{k}
		+
		\sum_{k=k^{*}+1}^{M} \Pbb\left( K_M=k \right) \frac{|\Acal|^n}{k} \nonumber\\
		&\le |\Acal|^n \cdot \Pbb\left( K_M \le k^* \right)
		+ \frac{|\Acal|^n}{k^*+1} \cdot \Pbb\left( K_M > k^* \right) \nonumber\\
		&\le |\Acal|^n \cdot \Pbb\left( K_M \le k^* \right)
		+ \frac{|\Acal|^n}{k^*},
	\end{align}
	for some $1 \le k^* \le M$.
	From~\eqref{eq:bounds-mean-KM} in Appendix~\ref{sec:appendix-proof-lemma-probability}, we have $\E\left[ \frac{K_M}{M} \right] \ge 1 - \frac{2^{nR}}{2|\Acal|^n}$; therefore, choosing $k^* \coloneqq k^*(n) \coloneqq M \left(1 - \frac{1}{n}\right)$ ensures that $\E\left[ \frac{K_M}{M} \right] - \frac{k^*}{M} \ge \frac{1}{n} - \frac{2^{nR}}{2|\Acal|^n}$, so that there exists an $n_0 \coloneqq n_0\left(|\Acal|,R\right)$ such that, for $n \ge n_0$, we have $\E\left[ \frac{K_M}{M} \right] - \frac{k^*}{M} \ge \frac{1}{n} - \frac{2^{nR}}{2|\Acal|^n} > 0$, in which case $k^* < \E[K_M]$ and we may apply Lemma~\ref{lemma:bound-probability-Km}, yielding
	\begin{align*}
		Q_{\theta,\tilde{P}_m}(n)
		\le 2^{n \zeta(n)} \left( \left( \frac{3}{\E\left[ \frac{K_M}{M} \right] -\frac{k^*}{M}} \right)^2
		+ \frac{|\Acal|^n}{k^*} \right).
	\end{align*}

	Let us study both terms inside the parenthesis. For $n \ge n_0$,
	\begin{align*}
		\frac{2}{n}\log \left( \frac{3}{\E\left[ \frac{K_M}{M} \right] -\frac{k^*}{M}} \right)
		\le \frac{2}{n} \log \left(\frac{3}{ \frac{1}{n} - \frac{2^{nR}}{2|\Acal|^n} }\right)
		\eqqcolon \lambda_1(n).
	\end{align*}
	Note that $\lim_{n\to\infty} \lambda_1(n) = 0$, with $\lambda_1(n) = O\left( (\log n)/n \right)$. Furthermore,
	\begin{align*}
		\frac{1}{n} \log \left(\frac{|\Acal|^n}{k^*(n)}\right)
		= \frac{1}{n} \log \left( \frac{|\Acal|^n}{2^{nR} \left( 1 - \frac{1}{n} \right)} \right)
		= \log|\Acal| - R + \underbrace{\frac{1}{n}\log \left(\frac{n}{n-1}\right)}_{\eqqcolon\lambda_2(n)}.
	\end{align*}
	Note that $\lim_{n\to\infty} \lambda_2(n) = 0$, with $\lambda_2(n) = O\left( 1/n^2 \right)$. We then have
	\begin{align}
		Q_{\theta,\tilde{P}_m}(n)
		&\le 2^{n \zeta(n)} \left( 2^{n\lambda_1(n)} + 2^{n\left( \log|\Acal| -R + \lambda_2(n) \right)} \right) \nonumber\\
		&\le 2^{n \zeta(n) } \left( 2 \cdot 2^{n \max\left\{ \lambda_1(n),\ \log|\Acal| - R + \lambda_2(n) \right\} } \right) \nonumber\\
		&= 2^{n \left( \max\left\{ \lambda_1(n),\ \log|\Acal| - R + \lambda_2(n) \right\} + \zeta(n) + \frac{1}{n} \right)}. \label{eq:proof-second-upper-bound}
	\end{align}

	The result of part (a) follows by combining \eqref{eq:proof-first-upper-bound} and \eqref{eq:proof-second-upper-bound}. The proof of part (b) is analogous.
\end{IEEEproof}

\section{Numerical Results} \label{sec:numerical-results}

In this section, we evaluate our decoding algorithms on simple examples with a binary alphabet $\Acal = \{0,1\}$. We use a modified version of a BCH code with parameters $n=63$ and $k = 51$ (where $M = 2^k$), as described next.
This code contains antipodal codeword pairs, i.e. pairs $(\xv,\bar{\xv})$ such that $\xv + \bar{\xv} = 1^n$. If a noise sequence $\zv$ is such that $\xv = \yv - \zv$ is a codeword, then subtracting its antipode $\bar{\zv}$ from the received sequence $\yv$ also results in a codeword, namely the antipodal one $\bar{\xv} = \yv - \bar{\zv}$. This is the case, for instance, of noise sequences $\zv = 0^n$ and $\bar{\zv} = 1^n$, which are the first two tested by the maximising and weighting strategies. In order to avoid the possible corresponding pair of codewords from being confused, we modify the code by keeping only one codeword in each antipodal pair (say, the one starting with a $0$), as in~\cite{miyamoto2024}. This reduces the effective code rate from $k/n$ to $(k-1)/n$.

We consider Markov additive channels of order $1$, i.e. $\Scal = \Acal = \{0,1\}$, parametrised by $\theta \coloneqq \left(\theta_0,\theta_1\right)$, where $\theta_s = P_{Z \given S}(1 \given s)$, for $s\in\{0,1\}$, and with fixed initial state $s_0=0$. The channel parameters can be organised in the stochastic matrix
\begin{equation*} \label{eq:stochastic-matrix-P}
	P =
	\begin{pmatrix}
		1-\theta_0 & \theta_0\\
		1-\theta_1 & \theta_1
	\end{pmatrix}.
\end{equation*}
We study two extreme symmetric regimes, namely, when $\theta_0 = \theta_1$ and $\theta_0 = 1- \theta_1$.

In our simulations, we consider both deterministic guessing~(DG) and randomised guessing~(RG) decoding with the weighting distribution. For the randomised version, a list decoding strategy is adopted: to decode each block, up to $L=20$ candidate codewords are selected by randomly drawing noise sequences, and the one with the highest metric (weighting distribution) is declared the decoded one.

These are compared to the following baseline strategies. First, deterministic guessing decoding with matched channel distribution~(GRAND)~\cite{duffy2019,an2022}---which would correspond to ML decoding, if the number of queries was unbounded\footnote{
	To limit the complexity, the number of queries is limited to $2^{n\left( \log|\Acal| - R \right)}$ in both deterministic and randomised guessing decoding, making the deterministic guessing decoder with matched distribution slightly deviate from true ML decoding. The abandonment strategy of~\cite{duffy2019} prescribed abandon after $\approx 2^{nH(P_{\theta})}$ queries to maintain the same asymptotic random-coding error exponent; but as this quantity depends on the channel, which we consider to be unknown to some of the decoders, we use a different threshold here.
}.
Second, a training-based scheme that uses the same $n=63$ binary channel uses, of which $8$~bits are used for a known training sequence, and $55$~bits for coded data from a punctured version of the original code. The decoder estimates the channel parameters $\hat{\theta}$ with the part of the received sequence corresponding to the training sequence, and applies deterministic guessing decoder on the coded sequence according to the distribution parametrised by $\hat{\theta}$.
Finally, a mismatched decoder that considers the channel to be a memoryless additive channel, with the stationary distribution of the noise sequence.

\begin{figure} 
	\centering
	\subfloat[$P = \left(\begin{smallmatrix}
		p & 1-p\\
		1-p & p
	\end{smallmatrix}\right)$\label{}]{%
		\includegraphics[width=0.45\linewidth]{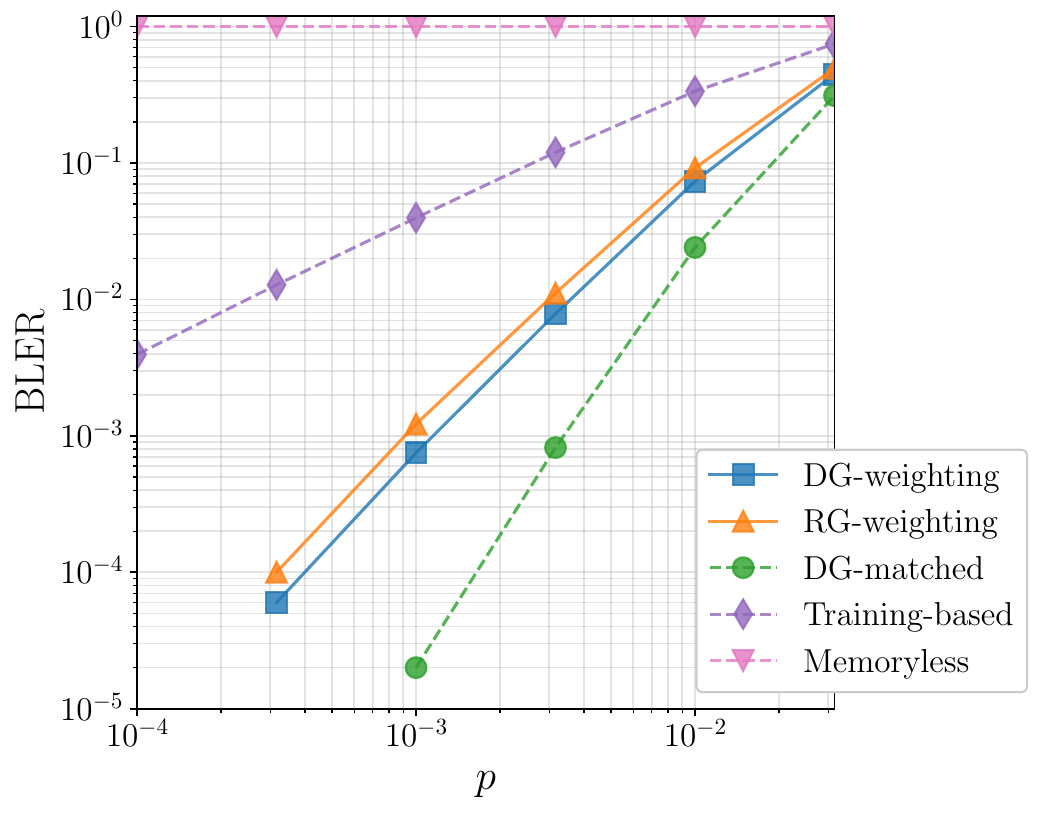}
	}
	\subfloat[$P = \left(\begin{smallmatrix}
		1-p & p\\
		p & 1-p
	\end{smallmatrix}\right)$\label{}]{%
		\includegraphics[width=0.45\linewidth]{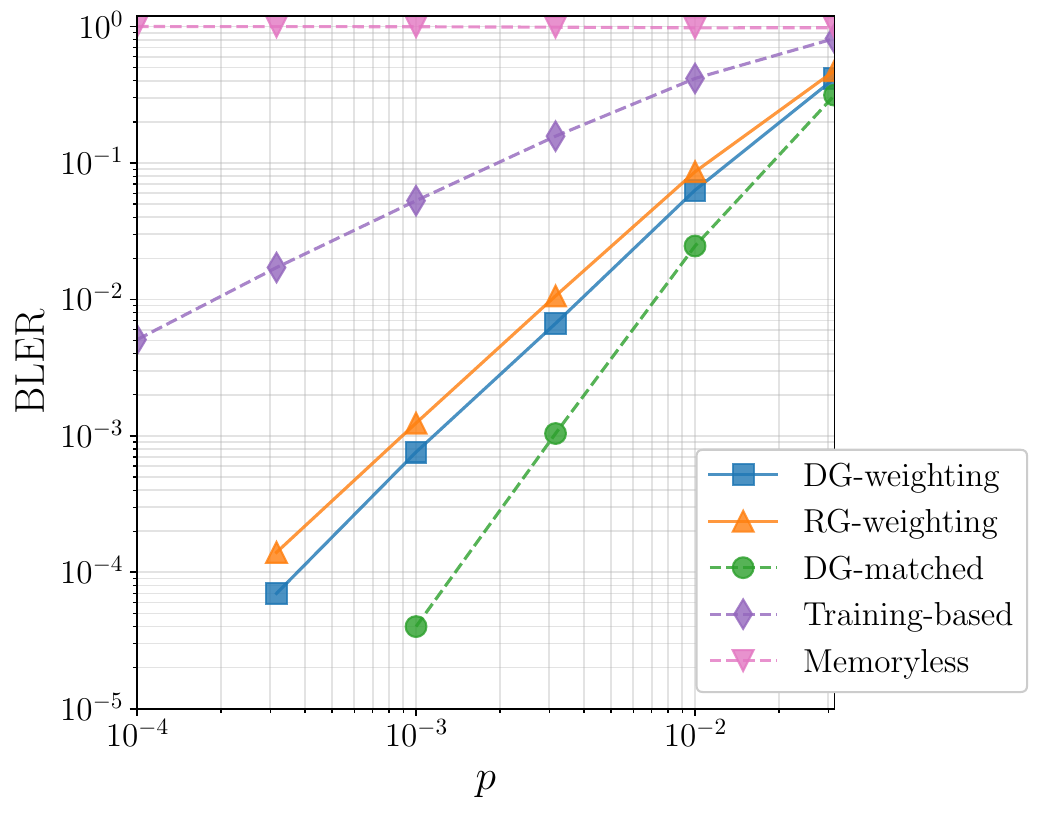}
	}
	\hfill
	\subfloat[$P = \left(\begin{smallmatrix}
		p & 1-p\\
		p & 1-p
	\end{smallmatrix}\right)$\label{}]{%
		\includegraphics[width=0.45\linewidth]{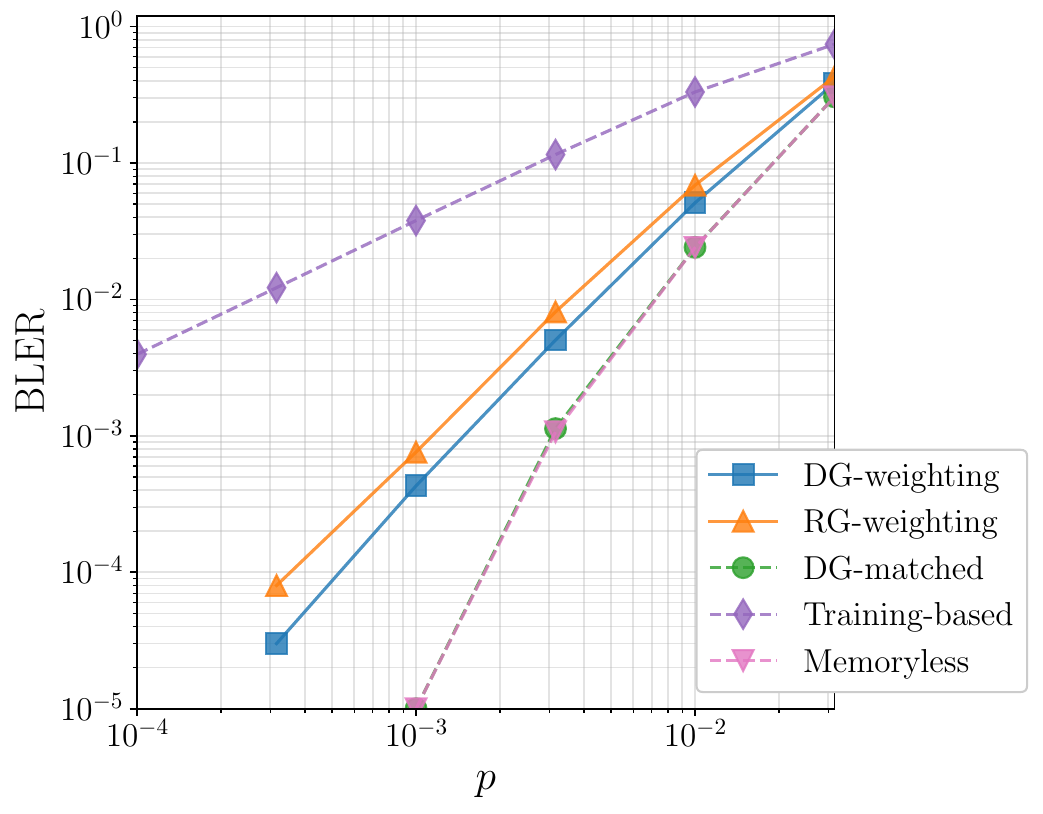}
	}
	\subfloat[$P = \left(\begin{smallmatrix}
		1-p & p\\
		1-p & p
	\end{smallmatrix}\right)$\label{}]{%
		\includegraphics[width=0.45\linewidth]{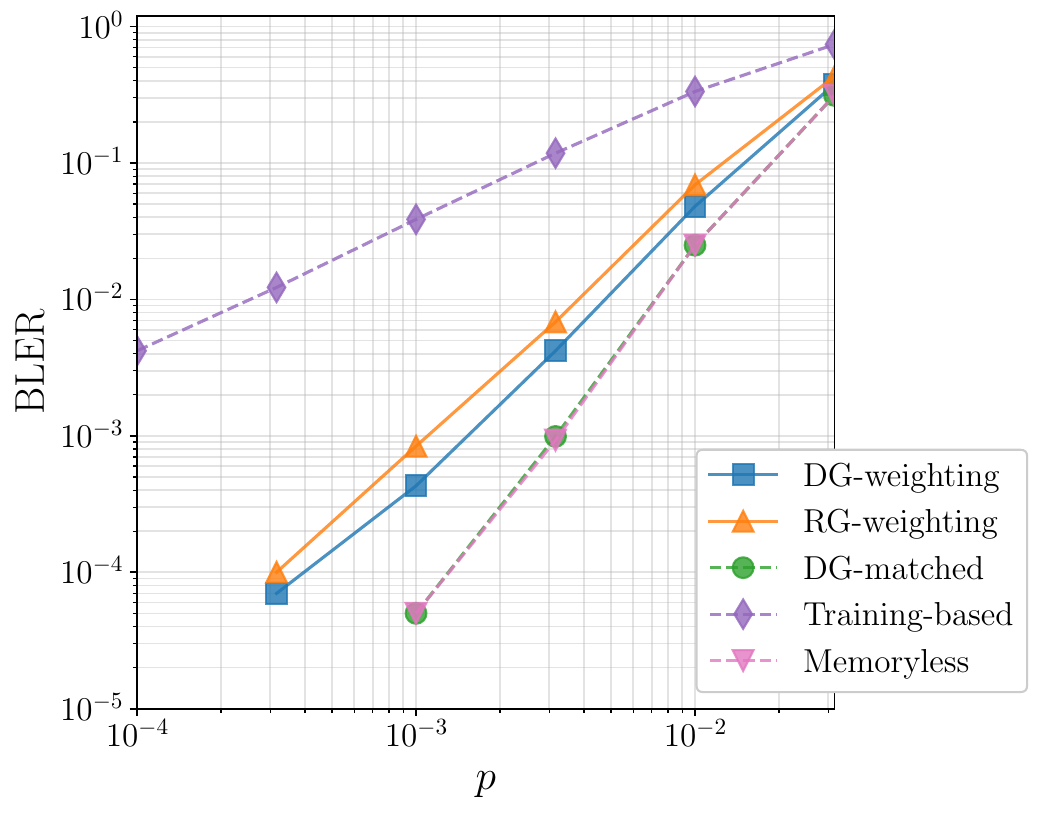}
	}
	\\
	\caption{Block error rate~(BLER) of different decoders in order-$1$ Markov additive channels with stochastic matrix~$P$.}
	\label{fig:bler} 
\end{figure}

Fig.~\ref{fig:bler} shows the block error rate~(BLER) of these strategies for different choices of stochastic matrix~$P$ parametrised by a single parameter $p\in\left[0,1\right]$. We note that the performances of randomised and deterministic guessing decoding are very close, and they consistently outperform the training-based strategy. In the first two channels, the memoryless decoder has error rate of one, which is explained by the fact that the stationary distribution in these channels is $\left( P_S(0), P_S(1) \right) = \left(1/2, 1/2\right)$, illustrating, in this extreme example, the effect that model mismatch can have on the decoding performance. The last two channels, on the other hand, are actually memoryless channels, so that the performance of the memoryless decoder matches that of the deterministic guessing decoder that is cognisant of the channel distribution.

We emphasise that the weighting and training-based decoders only know that the noise distribution is a Markov process of order 1, but they are unaware of the specific structure of the stochastic matrix (i.e. the symmetric regimes we study). Furthermore, decoding (and estimation of the channel, for the training-based decoder) is done independently for each codeword, in the sense that information from previously decoded codewords are not used to possibly improve the decoding (or estimation) of the current block.
This kind of strategy could be of interest in short packet length scenarios with low latency requirements, in which it is not possible to first perform a good estimation of the channel to be kept fixed for the rest of transmission.
Additionally, it could be of interest in slowly varying channels, in which the channel parameters are fixed during a block of length $n$, but may to change from one block to the other.

\begin{figure} 
	\centering
	\subfloat[$P = \left(\begin{smallmatrix}
		p & 1-p\\
		1-p & p
	\end{smallmatrix}\right)$\label{}]{%
		\includegraphics[width=0.45\linewidth]{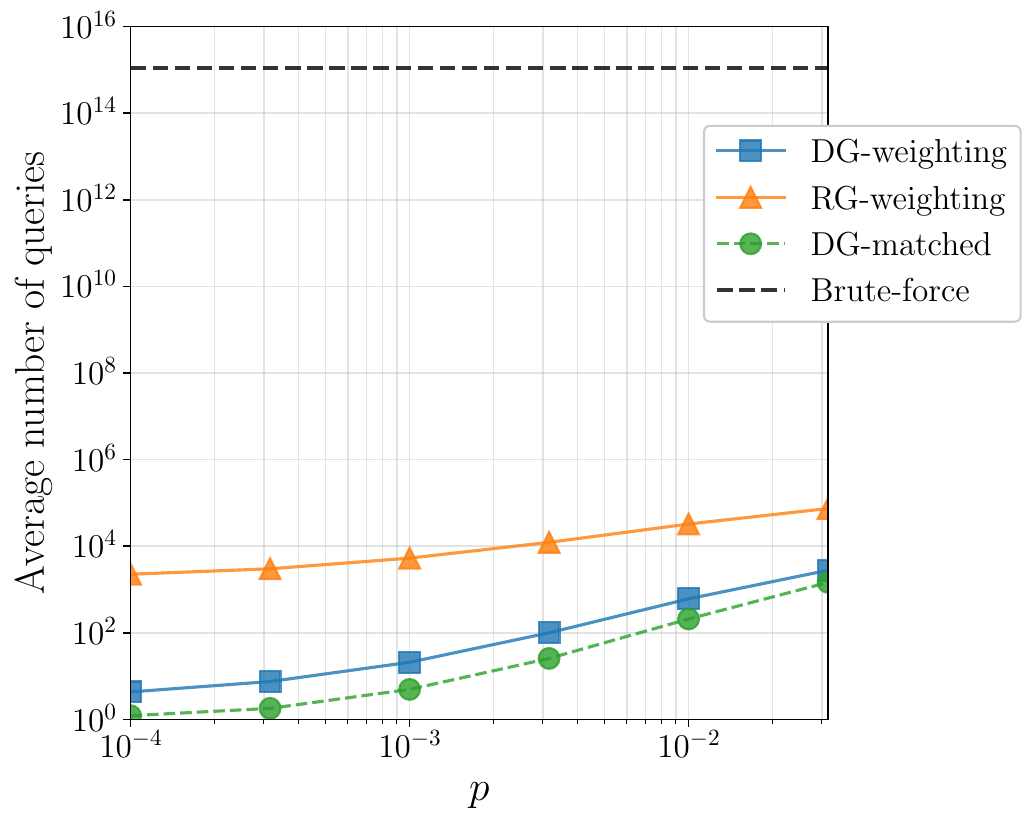}
	}
	\subfloat[$P = \left(\begin{smallmatrix}
		1-p & p\\
		p & 1-p
	\end{smallmatrix}\right)$\label{}]{%
		\includegraphics[width=0.45\linewidth]{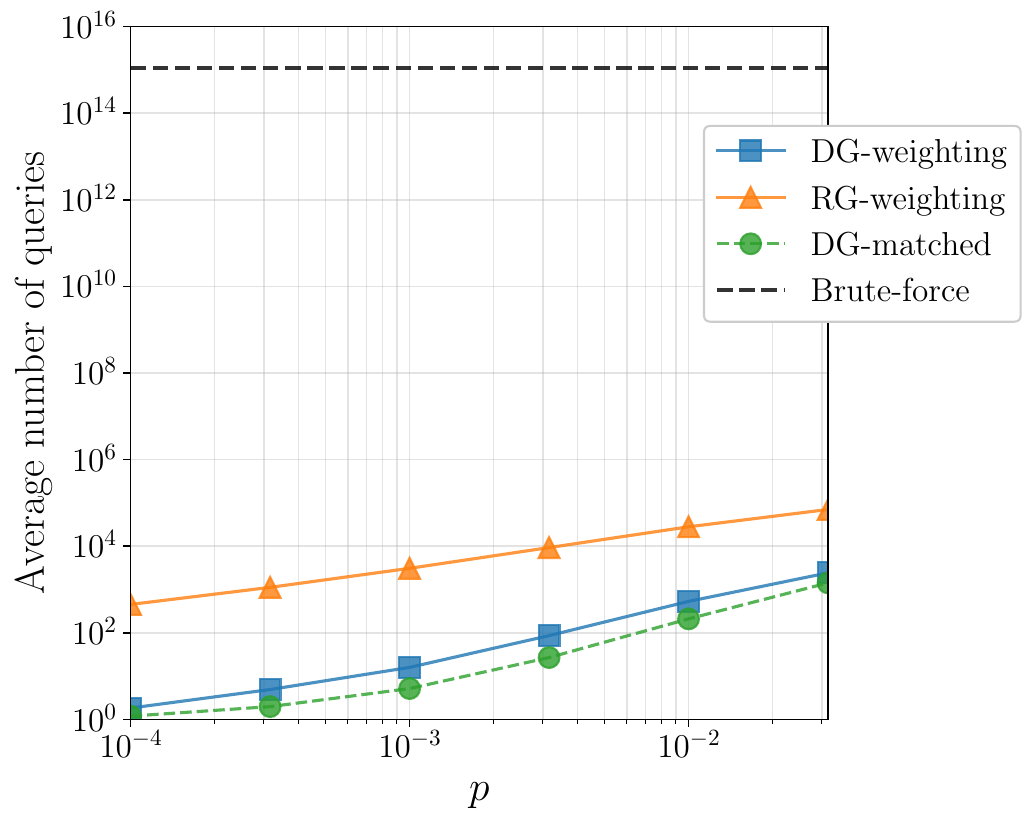}
	}
	\hfill
	\subfloat[$P = \left(\begin{smallmatrix}
		p & 1-p\\
		p & 1-p
	\end{smallmatrix}\right)$\label{}]{%
		\includegraphics[width=0.45\linewidth]{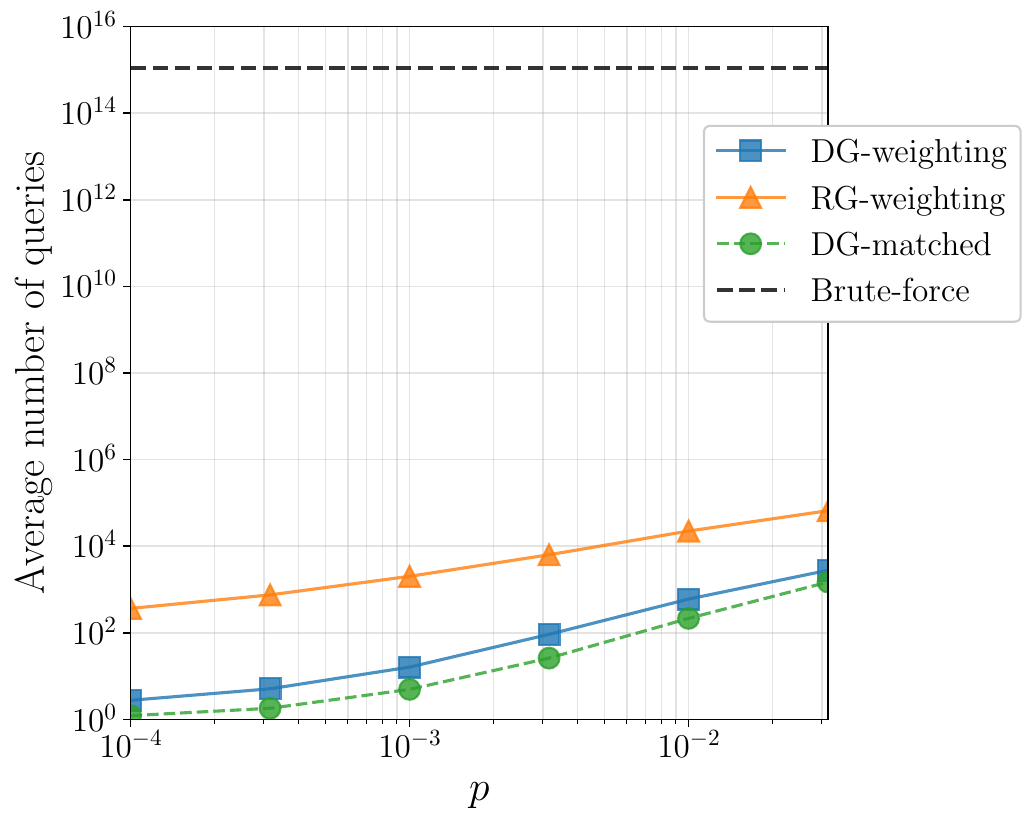}
	}
	\subfloat[$P = \left(\begin{smallmatrix}
		1-p & p\\
		1-p & p
	\end{smallmatrix}\right)$\label{}]{%
		\includegraphics[width=0.45\linewidth]{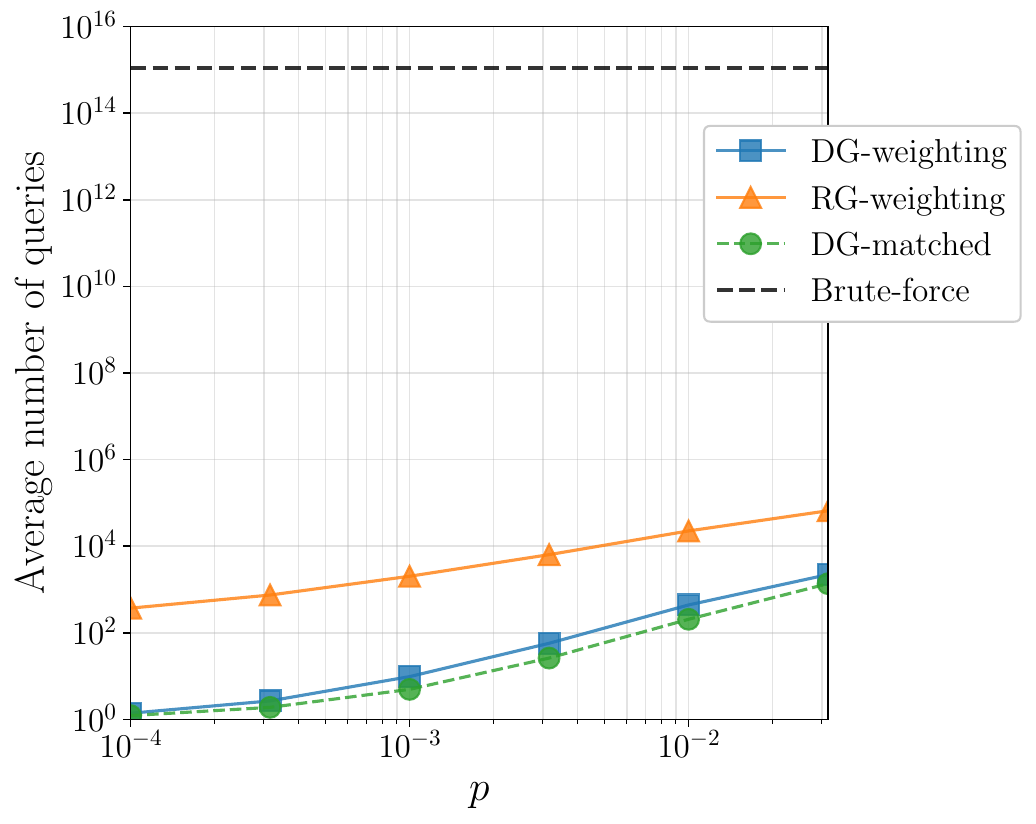}
	}
	\\
	\caption{Complexity of different decoders in order-$1$ Markov additive channels with stochastic matrix~$P$.}
	\label{fig:complexity}
\end{figure}

In Fig.~\ref{fig:complexity} we show the complexity as the average number of queries required by each strategy to find a codeword, in the same channels as before. For comparison, the codebook size is also shown, which corresponds to the number of evaluations needed in brute-force decoding (with ML or any other metric). We see that the average number of queries of guessing strategies is many orders of magnitude lower than that of the brute-force decoding.
While the number of queries needed by randomised guessing decoding is greater than that of deterministic guessing method, we emphasise that the cost of generating a randomised guess is usually lower than that of a deterministic guess, because it bypasses the need for evaluating each type, cf. Section~\ref{sec:randomised-guessing}. In particular, when the number of states (e.g. the memory of Markov process) increases, the complexity of deterministic guessing decoding may render it impractical, leaving randomised guessing decoding as the only feasible alternative.

\section{Conclusion} \label{sec:conclusion}

We have studied universal decoding over additive channels beyond memoryless distributions, considering noise sequences generated by the a general finite-state (unifilar) process.
The proposed techniques are based on deterministic noise guessing, in which noise sequences are queried in the order induced by a decoding metric, and randomised noise guessing, in which noise sequences are randomly drawn according to some probability distribution.
In both cases, we describe a low-complexity decoding scheme for which random-coding universality is proved and upper bounds on the complexity are established.
Numerical results show the feasibility and effectiveness of the proposed schemes, which outperform a training-based decoding strategy. This could lead to practical universal decoding, when the channel parameters are unknown.

\section*{Acknowledgments}
Discussions with M.~Pierre are gratefully acknowledged.

\appendices

\section{Proof of Lemma~\ref{lemma:bound-variational-characterisation}} \label{sec:appendix-proof-lemma-variational}

\begin{IEEEproof}
	An $n$-type does not necessarily define a stationary, irreducible Markov chain on $\Scal$, but a sequence with a circular convention does~\cite{davisson1981}. Specifically, given initial state $s_0 \in \Scal$ and sequence $\zv \in \Acal^n$, consider the associated sequence of states $\sv \in \Scal^n$, given by $s_i = f(z_i, s_{i-1})$. We now define the sequence $\tilde{\sv} \coloneqq s_1 \cdots s_n s_{n+1} \cdots s_{n+c-1} s_0 \in \Scal^{n+c}$ which is formed by appending to $\sv$ a suffix $s_{n+1} \cdots s_{n+c-1} s_0 \in \Scal^c$ that ends with the initial state $s_0$ and such that ${P_{\Sv}(\tilde{\sv} \given s_0) > 0}$. This is always possible because we suppose the Markov chain associated to $P_{\theta} \in \Fcal^{\star}$ to be irreducible. This suffix has length $c \coloneqq c(s_0,s_n)$ satisfying $0 \le c \le |\Scal|$. The associated symbol sequence is $\tilde{\zv} \in \Acal^{n+c}$, and we can now consider the counts $\tilde{a}_{\zv}(z,s) \coloneqq a_{\tilde{\zv}}(z,s) = \sum_{i=1}^{n+c} \1_{\{(z,s)\}}  (\tilde{z}_{i},\tilde{s}_{i-1})$ and $\tilde{a}_{\zv}(s) \coloneqq \sum_{z\in\Acal} \tilde{a}_{\zv}(z,s)$; as well as the corresponding distribution $\tilde{P}_{\zv}$, given by
	\begin{equation*}
		\tilde{P}_{\zv}\left(z,s\right) \coloneqq \frac{\tilde{a}_{\zv}(z,s)}{n}
	\end{equation*}
	and
	\begin{equation*}
		\tilde{P}_{\zv}\left(z \given s\right)
		\coloneqq
		\begin{cases}
			\frac{\tilde{a}_{\zv}(z,s)}{\tilde{a}_{\zv}(s)}, & \text{if } \tilde{a}_{\zv}(s) \neq 0,\\
			0, &\text{otherwise}.
		\end{cases}
	\end{equation*}
	
	This distribution defines an irreducible, stationary Markov chain~\cite[p.~434]{davisson1981}. Note that $a_{\zv}(z,s) \le \tilde{a}_{\zv}(z,s) \le a_{\zv}(z,s) + 1$ and $a_{\zv}(s) \le \tilde{a}_{\zv}(s) \le a_{\zv}(s) + 1$.
	We have
	\begin{align}
		&\hspace{-2.5em}\left( H(\hat{P}_{\zv}) - \frac{\alpha}{1-\alpha} D(\hat{P}_{\zv} \| P_{\theta}) \right)
		- \left( H(\tilde{P}_{\zv}) - \frac{\alpha}{1-\alpha} D(\tilde{P}_{\zv} \| P_{\theta}) \right) \nonumber\\
		&= \sum_{(z,s) \in \Acal\times\Scal} \hat{P}_{\zv}(z,s) \log \frac{P_{\theta}(z \given s)^{\alpha/(1-\alpha)}}{\hat{P}_{\zv}(z \given s)^{1/(1-\alpha)}}
		- \sum_{(z,s) \in \Acal\times\Scal} \tilde{P}_{\zv}(z,s) \log \frac{P_{\theta}(z \given s)^{\alpha/(1-\alpha)}}{\tilde{P}_{\zv}(z \given s)^{1/(1-\alpha)}} \nonumber\\
		&= \frac{1}{1-\alpha} \left( H(\hat{P}_{\zv}) - H(\tilde{P}_{\zv}) \right)
		+ \frac{\alpha}{1-\alpha} \sum_{(z,s) \in \Acal\times\Scal} \left( \frac{a_{\zv}(z,s)}{n} - \frac{\tilde{a}_{\zv}(z,s)}{n+c} \right) \log P_{\theta}(z \given s). \label{eq:appendix-1-initial}
	\end{align}

	We can bound the difference of entropies as
	\begin{align*}
		H(\hat{P}_{\zv}) - H(\tilde{P}_{\zv})
		&= \sum_{(z,s)\in\Acal\times\Scal} \frac{a_{\zv}(z,s)}{n} \log \frac{a_{\zv}(s)}{a_{\zv}(z,s)}
		- \sum_{(z,s)\in\Acal\times\Scal} \frac{\tilde{a}_{\zv}(z,s)}{n+c} \log \frac{\tilde{a}_{\zv}(s)}{\tilde{a}_{\zv}(z,s)}\\
		&\le \sum_{(z,s)\in\Acal\times\Scal} \frac{a_{\zv}(z,s)}{n} \log \frac{a_{\zv}(s)}{a_{\zv}(z,s)}
		- \sum_{(z,s)\in\Acal\times\Scal} \frac{a_{\zv}(z,s)}{n+c} \log \frac{a_{\zv}(s)}{a_{\zv}(z,s)+1}\\
		&= \left( 1 - \frac{n}{n+c} \right) \sum_{(z,s)\in\Acal\times\Scal} \frac{a_{\zv}(z,s)}{n} \log \frac{a_{\zv}(s)}{a_{\zv}(z,s)}
		+ \frac{1}{n} \sum_{(z,s)\in\Acal\times\Scal} a_{\zv}(z,s) \log \left( 1 + \frac{1}{a_{\zv}(z,s)} \right)\\
		&\le \left( \frac{c}{n+c} \right) H(\hat{P}_{\zv})
		+ \frac{|\Acal||\Scal| \log e}{n}\\
		&\le \frac{|\Scal| \log |\Acal|}{n}
		+ \frac{|\Acal||\Scal| \log e}{n},
	\end{align*}
	where in the second inequality we used the fact that $x \log \left( 1 + 1/x \right) \le \log e$, for $x\ge 0$.

	For the second term, we have
	\begin{align*}
		\sum_{(z,s) \in \Acal\times\Scal} \left( \frac{a_{\zv}(z,s)}{n} - \frac{\tilde{a}_{\zv}(z,s)}{n+c} \right) \log P_{\theta}(z \given s)
		&\le \sum_{(z,s) \in \Acal\times\Scal} \left( \frac{a_{\zv}(z,s)}{n} - \frac{{a}_{\zv}(z,s)+1}{n+c} \right) \log P_{\theta}(z \given s)\\
		&\le -\frac{1}{n} \sum_{(z,s) \in \Acal\times\Scal} \log {P_{\theta}(z \given s)}\\
		&\le -\frac{|\Acal||\Scal|}{n} \log \left( \min_{(z,s)\in\Acal\times\Scal \colon P_{\theta}(z \given s) > 0} P_{\theta}(z \given s) \right).
	\end{align*}
	For convenience, denote $P^*_{\theta} \coloneqq \min \left\{ P_{\theta}(z \given s) \colon P_{\theta}(z \given s) > 0\right\}$. We can restrict the minimum to pairs $(z,s)$ for which $P_{\theta}(z \given s)$ is strictly positive, because both $\hat{P}_{\zv}$ and $\tilde{P}_{\zv}$ are dominated by $P_{\theta}$, so the original sum in \eqref{eq:appendix-1-initial} is also restricted to this set.

	Putting everything together gives us
	\begin{align*}
		&\hspace{-2em}H(\hat{P}_{\zv}) - \frac{\alpha}{1-\alpha} D(\hat{P}_{\zv} \| P_{\theta})\\ 
		&\le H(\tilde{P}_{\zv}) - \frac{\alpha}{1-\alpha} D(\tilde{P}_{\zv} \| P_{\theta}) 
		+ \frac{|\Scal|}{n(1-\alpha)} \left( |\Acal| \log e + \log |\Acal| - \alpha |\Acal| \log P^*_\theta \right)
		\\
		&\le \max_{Q \in \Fcal^{\star}} \left( H(Q) - \frac{\alpha}{1-\alpha} D(Q \| P_{\theta}) \right)
		+ \frac{|\Scal|}{n(1-\alpha)} \left( |\Acal| \log e + \log |\Acal| - \alpha |\Acal| \log P^*_\theta \right)\\
		&= H_{\alpha}(P_{\theta})
		+ \frac{|\Scal|}{n(1-\alpha)} \left( |\Acal| \log e + \log |\Acal| - \alpha |\Acal| \log P^*_\theta \right),
	\end{align*}
	where the second inequality is due to the fact that $\tilde{P}_{\zv} \in \Fcal^{\star}$, and the last equality, due to Lemma~\ref{lemma:renyi-entropy}. This holds for any sequence $\zv \in \Acal^n$, and in particular for a sequence corresponding to the maximising type on the left-hand side of \eqref{eq:renyi-entropy-variational-bound}, which concludes the proof.
	
\end{IEEEproof}

\section{Proof of Lemma~\ref{lemma:expectation-B}} \label{sec:appendix-proof-lemma-expectation}

\begin{IEEEproof}
	To study the statistics of $K = \min_{m' \neq 1} K_{m'}$, we first write
	\begin{align*}
			\Pbb \left( K \ge k \right)
			&= \Pbb \left( \min_{m'\in\{2,\dots,M\}} K_{m'} \ge k \right)\\
			&= \Pbb \left( \bigcap_{m'=2}^{M} \left\{ K_{m'} \ge k \right\} \right)\\
			&= \left(\frac{|\Acal|^n - k + 1}{|\Acal|^n}\right)^{M-1},
		\end{align*}
	so that
	\begin{align*}
			\Pbb \left( K = k \right)
			&= \Pbb \left( K \ge k \right) - \Pbb \left( K \ge k+1 \right)\\
			&= \frac{\left( |\Acal|^n - k +1 \right)^{M-1} - \left( |\Acal|^n - k \right)^{M-1} }{|\Acal|^{n(M-1)}}.
		\end{align*}
	The expectation is then
	\begin{align*}
			\E[K]
			= \sum_{k=1}^{|\Acal|^n} k \cdot \Pbb(K=k)
			= \sum_{k=1}^{|\Acal|^n} {\left(\frac{k}{|\Acal^n|}\right)}^{M-1}.
		\end{align*}
	
	For the lower bound, we upper bound the integral $\int_0^{1} x^{M-1}~\d x$ by a right Riemann sum with interval length $\Delta = 1/|\Acal|^n$:
	\begin{align*}
		\sum_{k=1}^{|\Acal|^n} \left(\frac{k}{|\Acal|^n}\right)^{M-1} \frac{1}{|\Acal^n|}
		\ge \int_0^1 x^{M-1}~\d x
		= \frac{1}{M},
	\end{align*}
	which yields $\E[K] \ge \frac{|\Acal|^n}{M} = 2^{n\left( \log|\Acal| - R \right)}$.
	For the upper bound, we similarly lower bound the same integral by a left Riemann sum:
	\begin{align*}
		\sum_{k=1}^{|\Acal|^n} \left(\frac{k-1}{|\Acal|^n}\right)^{M-1} \frac{1}{|\Acal^n|}
		\le \int_0^1 x^{M-1}~\d x
		= \frac{1}{M},
	\end{align*}
	yielding $		\E\left[ K \right]
	\le \frac{|\Acal|^n}{M} + 1
	< 2 \frac{|\Acal|^n}{M}
	= 2^{n\left( \log|\Acal| - R \right)}$.
\end{IEEEproof}

\section{Proof of Lemma~\ref{lemma:bound-probability-Km}} \label{sec:appendix-proof-lemma-probability}

	Recall that $K_M = K_M(\Ccal)$ is the number of distinct codewords in the code $\Ccal = \left( \xv_1, \dots, \xv_M \right)$, which are drawn with independent, uniform distribution in $|\Acal|^n$. We will first compute the mean and variance of $K_M$ with an inductive argument on $M$.
	In the following, it will be useful to define the random variable $B_{M+1} \coloneqq B_{M+1}\left(\xv_1,\dots,\xv_{M+1}\right) \coloneqq 1 - \1_{\left\{ \xv_1, \dots, \xv_M \right\}}(\xv_{M+1})$ that indicates if the $(M+1)$-th sequence is distinct from the previous $\left( \xv_1, \dots, \xv_M \right)$. Its expectation is
	\begin{align*}
		\E\left[ B_{M+1} \right]
		&= \sum_{k=1}^{M} \Pbb\left( K_M = k \right) \E\left[ B_{M+1} \mgiven K_M = k \right]\\
		&= \sum_{k=1}^{M} \Pbb\left( K_M = k \right) \left(\frac{|\Acal|^n-k}{|\Acal|^n}\right)\\
		&= 1 - \frac{a_M}{|\Acal|^n}.
	\end{align*}
	
	Denote the mean $a_M \coloneqq \E\left[ K_M \right]$. Note that $a_{M+1} - a_{M} = \E\left[ K_{M+1} - K_M \right] = \E\left[ B_{M+1} \right]$. Hence, $a_{M+1} = \left( 1 - \frac{1}{|\Acal|^n} \right) a_M + 1$. Together with $a_1 = 1$, one can recursively show that
	\begin{align} \label{eq:mean-aM}
		\E\left[ K_M \right]
		= |\Acal|^n \left[ 1 - \left( 1 - \frac{1}{|\Acal|^n} \right)^{M} \right].
	\end{align}
	
	For the variance $\Var\left( K_M \right) = \E\left[ \left(K_M\right)^2 \right] - a_M^2$, denote $s_M \coloneqq \E\left[ \left(K_M\right)^2 \right]$ and note that
	\begin{align*}
		s_{M+1}
		= \E\left[ \left( K_M + B_{M+1} \right)^2 \right]
		= \E\left[ \left(K_M\right)^2 \right] + \E\left[ \left(B_{M+1}\right)^2 \right] + 2\E\left[ K_M B_{M+1} \right].
	\end{align*}
	We have $\E\left[ \left(B_{M+1}\right)^2 \right] = \E\left[ B_{M+1} \right]$, since $B_{M+1} \in \{0,1\}$. Moreover,
	\begin{align*}
		\E\left[ K_M B_{M+1} \right]
		&= \Pbb\left( B_{M+1} = 1 \right) \E\left[ K_M B_{M+1} \mgiven B_{M+1} = 1 \right]\\
		&\quad+ \Pbb\left( B_{M+1} = 0 \right) \E\left[ K_M B_{M+1} \mgiven B_{M+1} = 0 \right]\\
		&= \Pbb\left( B_{M+1} = 1 \right) \E\left[ K_M \right]\\
		&= \left( 1 - \frac{a_M}{|\Acal|^n} \right) a_M,
	\end{align*}
	where we used that $\E\left[ B_{M+1} \right] = \Pbb\left( B_{M+1} = 1 \right)$. We thus have
	\begin{align*}
		s_{M+1}
		&= s_M + \left( 2a_M + 1 \right) \left(1 - \frac{a_M}{|\Acal|^n}\right).
	\end{align*}
	Together with $s_1=1$ and the expression~\eqref{eq:mean-aM} for $a_M$, one can recursively show, after some calculation, that
	\begin{equation*} \label{eq:second-moment-sM}
		s_M = \left( 2|\Acal|^{2n} + |\Acal|^n \right) \left[ 1 - \left(1-\frac{1}{|\Acal|^n}\right)^M \right]
		- \frac{2|\Acal|^{3n}}{2|\Acal|^n-1}\left[ 1 - \left(1-\frac{1}{|\Acal|^n}\right)^{2M} \right].
	\end{equation*}
	Replacing in the expression for the variance and simplifying, we get
	\begin{align} \label{eq:var-var-Km}
		\Var\left( K_M \right)
		&= \frac{|\Acal|^n}{2 |\Acal|^n-1}
		\left[ 1-\left(1-\frac{1}{|\Acal|^n}\right)^M \right]
		\left[ |\Acal|^n \left(1-\left(1-\frac{1}{|\Acal|^n}\right)^M\right) - 1 \right].
	\end{align}

	Writing the Taylor expansions around the origin with Lagrange remainder, we get
	${-x -\frac{x^2}{(1-x)^2}} \le \ln(1-x) \le -x$, for $x\in\left[0,1\right]$, and $1 - x \le e^{-x} \le 1 - x + \frac{x^2}{2}$, for $x\in\R$.
	By writing $\left( 1 - \frac{1}{|\Acal|^n} \right)^{2^{nR}} = \exp\left( 2^{nR} \, \ln \left( 1 - \frac{1}{|\Acal|^n} \right) \right)$ and applying the previous bounds, one has the estimation $1 -\frac{2^{nR}}{|\Acal|^n} - \frac{2^{nR}}{\left( |\Acal|^n-1 \right)^2} \le \left( 1 - \frac{1}{|\Acal|^n} \right)^{2^{nR}}
	\le 1 - \frac{2^{nR}}{|\Acal|^n} + \frac{1}{2}\left( \frac{2^{nR}}{|\Acal|^n} \right)^2$,
	which, applied to~\eqref{eq:mean-aM} and \eqref{eq:var-var-Km}, yields
	\begin{equation} \label{eq:bounds-mean-KM}
		1 - \frac{2^{nR}}{2|\Acal|^n}
		\le \E\left[\frac{K_M}{M}\right]
		\le 1 + \frac{|\Acal|^n}{\left( |\Acal|^n-1 \right)^2},
	\end{equation}
	and
	\begin{align} \label{eq:bounds-variance-KM}
		&\left(\frac{1}{2 |\Acal|^n-1}\right)
		\left( 1 - \frac{2^{nR}}{2 |\Acal|^n} \right)
		\left( 1 - \frac{2^{nR}}{2 |\Acal|^n} - \frac{1}{2^{nR}} \right) \nonumber\\
		&\hspace{2.5em}\le \Var\left(\frac{K_M}{M}\right) \nonumber\\
		&\hspace{2.5em}\le  \left(\frac{1}{2 |\Acal|^n-1}\right)
		\left( 1 + \frac{|\Acal|^n}{\left( |\Acal|^n-1 \right)^2} \right)
		\left( 1 + \frac{|\Acal|^n}{\left( |\Acal|^n-1 \right)^2} - \frac{1}{2^{nR}} \right).
	\end{align}

\begin{remark}
	Note that $\lim_{n\to\infty} \E\left[\frac{K_M}{M}\right] = 1$ and $\lim_{n\to\infty} \Var\left(\frac{K_M}{M}\right) = 0$, showing that the normalised random variable $\frac{K_M}{M} \in \left[0,1\right]$ concentrates around $1$.
\end{remark}

\begin{IEEEproof}[Proof of Lemma~\ref{lemma:bound-probability-Km}]
	Recall that we suppose $1 \le k^* < \E[K_M]$. Using Chebyshev's inequality, we then have
	\begin{align*}
		\Pbb\left( K_M \le k^{*} \right)
		&\le \Pbb\left( \left| K_M - \E\left[K_M\right] \right| \ge \E\left[ K_M \right] - k^* \right)\\
		&\le \frac{\Var\left(K_M\right)}{\left( \E\left[ K_M \right] - k^* \right)^2}\\
		&= \frac{\Var\left(\frac{K_M}{M}\right)}{\left( \E\left[\frac{K_M}{M}\right] - \frac{k^*}{M} \right)^{2}}.
	\end{align*}

	For the numerator, we have, thanks to \eqref{eq:bounds-variance-KM},
	\begin{align*}
		\Var\left( \frac{K_M}{M} \right)
		\le \frac{1}{2|\Acal|^n - 1} \left( 1 + \frac{|\Acal|^n}{\left( |\Acal|^n - 1 \right)^2} \right)^2
		\le \frac{9}{|\Acal|^n},
	\end{align*}
	where we used that $\frac{1}{2 |\Acal|^n-1} \le \frac{1}{|\Acal|^n}$ and $\frac{|\Acal|^n}{\left( |\Acal|^n-1 \right)^2} \le 2$. Therefore,
	\begin{align*}
		\Pbb\left( K_M \le k^{*} \right)
		\le \frac{1}{|\Acal|^n} \frac{9}{\left( \E\left[\frac{K_M}{M}\right] - \frac{k^*}{M} \right)^{2}}
		= \frac{1}{|\Acal|^n} \left( \frac{3 \cdot 2^{nR}}{\E\left[ K_M \right]-k^*} \right)^2.
	\end{align*}
\end{IEEEproof}

\bibliographystyle{IEEEtran}

\end{document}